\documentclass[fleqn,usenatbib]{mnras}
\usepackage{newtxtext,newtxmath}

\usepackage[T1]{fontenc}
\usepackage{chngcntr}
\usepackage{appendix}

\DeclareRobustCommand{\VAN}[3]{#2}
\let\VANthebibliography\thebibliography
\def\thebibliography{\DeclareRobustCommand{\VAN}[3]{##3}\VANthebibliography}

\usepackage{threeparttable}
\usepackage{graphicx}	
\usepackage{amsmath}	
\usepackage{amssymb}	

\newenvironment{rcases}
  {\left.\begin{aligned}}
  {\end{aligned}\right\rbrace}

\newcommand{\HII}{\mathrm{H}\,\textsc{\large{ii}}}

\graphicspath{{plots/}}

\title[Chemistry of 6.7 GHz maser sources]{Chemical environments of 6.7 GHz methanol maser sources}

\author[Paulson \& Pandian]{
Sonu Tabitha Paulson,$^{1}$\thanks{E-mail: sonutabitha.15@res.iist.ac.in}
Jagadheep D. Pandian$^{1}$\thanks{E-mail: jagadheep@iist.ac.in }
\\
$^{1}$Department of Earth and Space Sciences, Indian Institute of Space Science \& Technology, Thiruvananthapuram 695547, Kerala, India\\
}

\date{Accepted 2021 November 2. Received 2021 October 29; in original form 2021 March 19}

\pubyear{2021}

\begin{document}
\label{firstpage}
\pagerange{\pageref{firstpage}--\pageref{lastpage}}
\maketitle

\begin{abstract}
6.7~GHz methanol masers are the brightest of class II methanol masers that are regarded as excellent signposts in the formation of young massive stars.
 We present here a molecular line study of 68 6.7~GHz methanol maser hosts chosen from the MMB catalogue, that have MALT90 data available. We performed (1) pixel-by-pixel study of 9 methanol maser sources that have high signal-to-noise ratio and (2) statistical study taking into account the entire 68 sources. We estimated the molecular column densities and abundances of N$_2$H$^+$(1-0), HCO$^+$(1-0), HCN(1-0) and HNC(1-0) lines. The derived abundances are found to be in congruence with the typical values found towards high mass star forming regions. We derived the column density and abundance ratios between these molecular species as an attempt to unveil the evolutionary stage of methanol maser sources. We found the column density and abundance ratio of HCN to HNC to increase and that of N$_2$H$^+$ to HCO$^+$ to decline with source evolution, as suggested by the chemical models. The HCN/HNC, N$_2$H$^+$/HCO$^+$, HNC/HCO$^+$ and N$_2$H$^+$/HNC ratios of the methanol maser sources are consistent with them being at a later evolutionary state than quiescent phase and possibly protostellar phase, but at an earlier stage than $\HII$ regions and PDRs. 
\end{abstract}


\section{INTRODUCTION}\label{S1}
High mass stars play a cardinal role in the evolution of galaxies by enriching the interstellar medium through stellar winds and supernovae. Understanding the formation of such stars in the Universe is therefore crucial for constraining the models of galaxy formation and evolution. In spite of this, high mass star formation still remains an elusive phenomenon and requires detailed exploration. We can attribute this to rapid formation which takes place while they are deeply embedded in dense clumps. This in itself introduces challenges in understanding their early formation phases, in addition to rarity and large distance scales.

One of the methods to probe the early evolutionary phases of high mass star formation is to make use of interstellar masers, among which class II methanol masers (the 6.7 GHz line being the brightest) are seen almost exclusively towards massive star forming regions \citep[e.g.,][and references therein]{menten1992vlbi,xu2008high,breen2013confirmation}. There is strong evidence that 6.7~GHz methanol masers trace a very early stage of massive star formation \citep[e.g.,][and references therein]{minier2001vlbi,ellingsen2006methanol,pandian2010spectral,billington2019atlasgal}. Since the chemical composition in star forming environments is highly sensitive to the physical conditions \citep{gerner2014chemical}, it is of interest to examine whether the chemical properties of the methanol maser hosts are in accordance with them tracing an early phase of massive star formation. Although there have been many studies that focus on the chemistry of high mass star forming regions \citep[e.g.][]{sanhueza2012chemistry,hoq2013chemical,miettinen2014malt90,zhang2016global}, chemical processes surrounding 6.7 GHz methanol maser hosts are not much discussed yet. For example, \citet{saral2018malt90} investigated the physical and chemical properties of 30 high mass star forming clumps, the chemistry of which was traced using four molecular species: N$_2$H$^+$, HCO$^+$, HCN and HNC. This was used to classify the clumps into protostellar candidates, young stellar objects (YSOs), and massive star forming regions (MSF), the latter containing $\HII$ regions, radio bright sources and methanol masers. 

The MALT90 survey \citep{jackson2013malt90} carried out a comprehensive study of molecular emission towards 3246 high-mass clumps detected in the ATLASGAL survey. It was found that ratios of integrated line intensities of several molecules such as HCO$^+$/HNC and HCN/HNC showed systematic variation with the evolutionary stage of the source \citep{rathborne2016molecular}. A similar study targeting a much wider range of molecules towards $\sim$600 high mass clumps, was performed by \citet{urquhart2019atlasgal}. They also discovered that several line ratios were probes of the evolutionary state of the source. However, a direct comparison to chemical models is not possible since the studies of both \citet{rathborne2016molecular} and \citet{urquhart2019atlasgal} focused on integrated line intensities rather than column densities. Since the column density shows additional dependence on the excitation temperature and the optical depth, one needs to carry out radiative transfer modelling of the molecular spectra in order to compare the results with chemical models. Since all the studies focused on the evolutionary state in the context of massive star formation, there is a lack of proper knowledge on the chemistry of methanol maser sources alone.

In \citet{paulson2020probing} (hereafter Paper~I), we analysed the evolutionary stage of 320 sources associated with 6.7 GHz methanol masers using spectral energy distributions of dust continuum maps. We concluded that while a majority of our sources are more evolved as compared to infrared dark clouds, they are still in an early evolutionary stage, with the mass to luminosity ratio being consistent with the sources being in accretion phase. In this paper, we use the MALT90 data to conduct molecular studies of 68 6.7~GHz methanol maser sources and investigate whether the evolutionary phase of methanol maser hosts inferred from chemical signatures aligns with results in Paper~I.

The structure of the paper is as follows. In Section~\ref{sec2}, we discuss the source selection and data analysis. The results of our study are presented in Section~\ref{sec3}. A detailed discussion of the results in the context of the evolutionary phase of 6.7 GHz methanol maser hosts is presented in Section~\ref{sec4}, and we summarize our results in Section~\ref{sec5}.

\section{SOURCE SELECTION AND DATA ANALYSIS}\label{sec2}
\subsection{MALT90 Data}
 \begin{table*}
\caption{Spectral lines in MALT90 survey}
         \label{molecules}
\begin{center}
         \centering
  \begin{tabular}{ccc}
    \hline
    \hline

     Transition & {Frequency (MHz) }  &{Tracer}    \\
    \hline
    HCO$^{+}$(1–0) &89188.526 &Density; Kinematics\\
    H$^{13}$CO$^{+}$(1–0) &86754.330
&Optical depth, Column density, V$_{LSR}$\\	
    N$_{2}$H$^{+}$(1–0) &93173.772 &Density, chemically robust \\
    HCN (1–0) &88631.847 &Density\\
    HNC(1–0) &90663.572 &Density; Cold chemistry\\ 
    $^{13}$CS (2–1) &92494.303 &Optical depth, Column density, V$_{LSR}$ \\
    CH$_3$CN 5(0)–4(0) &91987.086 &Hot core \\
    HC$_3$N (10–9) &90978.989 &Hot core \\
    $^{13}$C${^34}$S (2–1) &90926.036 &Optical depth, Column density, V$_{LSR}$ \\
    HC$^{13}$CCN (10–9) &90593.059 &Hot core \\
    HNCO 4(1,3)–3(1,2) &88239.027 &Hot core \\
    HNCO 4(0,4)–3(0,3) &87925.238 &Hot core \\
    C$_2$H (1–0) 3/2–1/2 &87316.925 &Photodissociation region \\
    HN$^{13}$C (1–0) &87090.859 &Optical depth, Column density, V$_{LSR}$ \\
    SiO (1–0) &86847.010 &Shock/outflow \\
 
    \hline
  \end{tabular}
  \end{center}
\end{table*}

The source sample of the present paper was chosen among the sources studied in Paper~I, that have MALT90 data available. The MALT90 survey aims at characterizing the physical and chemical properties of massive star formation in our Galaxy. The survey is targeted towards 2014 compact sources detected in the ATLASGAL survey \citep{schuller2009atlasgal} covering Galactic longitude ranges $300^{\circ}<l<357^{\circ}$ and $3^{\circ}<l<20^{\circ}$, with the Mopra Spectrometer (MOPS)\footnote{The University of New South Wales Digital Filter Bank used for the observations with the Mopra Telescope was provided with support from the Australian Research Council.} arrayed on the Mopra 22 m telescope. The survey obtained $3' \times 3'$ maps around each source covering a total of 3264 high-mass clumps. MALT90 has mapped 16 molecular lines simultaneously at frequencies near 90 GHz with a velocity resolution of 0.11 km s$^{-1}$. The 16 spectral lines and their rest frequencies are shown in Table~\ref{molecules} \citep{jackson2013malt90}. The beam size of Mopra is 38$''$ at 86 GHz, with a main beam efficiency of 0.49 \citep{ladd2005beam}. Among the 320 sources listed in Paper~I, 138 of them were found to be included in the MALT90 survey. We have further narrowed down our study to sources that have strong detections (S/N~$\geq3$ in the moment zero map) in  HCO$^{+}$(1-0), HCN(1-0), HNC(1-0) and  N$_{2}$H$^{+}$(1-0). These molecular transitions are considered the brightest of the transitions covered by the MALT90 survey and have shown to be good tracers of density. In order to verify the association of the methanol masers with the MALT90 clump, we have compared the peak velocity of maser emission with that of molecular emission from the clump. We find that the velocity offsets are less than 10 km s$^{-1}$, with the mean offset being 3 km s$^{-1}$, which is similar to that observed in earlier studies \citep[e.g.][]{billington2019atlasgal}. We investigated the spatial matching of methanol maser sites and MALT90 clumps by analysing the differences in their positions, obtained from their catalogues. If the angular offset was found to be less than half of the MALT90 beam size, the maser is considered to be physically associated with the clump. If more than one clump is found to be matched with a particular maser, we chose the clump having the least angular offset as the physically associated one. The angular offsets between maser sites and clumps for the 68 MM sources were typically less than 6 arcseconds. This suggests that all the methanol masers in this study are indeed physically associated with the MALT90 clumps.

The molecular line modelling was performed in CASSIS \citep{Caux2011}, under the assumption of local thermodynamic equilibrium (LTE). CASSIS gives the best fit estimates for molecular column density, excitation temperature, line width and LSR velocity based on the initial parameters provided for the same. The values of initial parameters are obtained by fitting the spectra using the CLASS program of the GILDAS\footnote{http://www.iram.fr/IRAMFR/GILDAS} software package. Prior to the fitting, the spectrum of each pixel in the data cube is extracted and data were converted from antenna temperatures to main beam temperatures by dividing it by main beam efficiency ($\eta=0.49$). For the molecules that exhibit hyperfine components (HCN and N$_{2}$H$^{+}$), fitting is performed using method HFS (hyperfine-structure fit). For the molecules that do not have hyperfine satellites (HCO$^{+}$ and HNC), method GAUSS~\footnote{A detailed description of method HFS and method GAUSS can be found here, http://www.iram.fr/IRAMFR/GILDAS/doc/html/classhtml/class.html} (Gaussian fit) is employed. The fit performed using method HFS gives the optical depth in addition to the LSR velocity and FWHM of the spectral line, while the method GAUSS gives an estimate of only the latter two parameters. We estimate the excitation temperature of the line using the antenna equation
\begin{equation}
T_{mb}= f[J(T_{ex})-J(T_{bg})] (1-e^{-\tau_{\nu}}) \label{eq:tex}
\end{equation}
where $T_{mb}$ is the main beam temperature, $f$ is the filling factor, $\tau_{\nu}$ is the optical depth of the line, $T_{bg}$ is the background temperature, and $J(T)$ is defined by 
\begin{equation}
J(T) = \dfrac{h\nu}{k} \dfrac{1}{e^{h\nu/kT}-1}
\end{equation}

\begin{table*}
\begin{tabular}{lllll}
MMB names      & ATLASGAL names     & MMB names          & ATLASGAL names     &  \\\hline\hline
G006.189-0.358 & AGAL006.188-00.357 & G338.497+0.207     & AGAL338.497+00.207 &  \\
G010.724-0.334 & AGAL010.724-00.332 & G338.566+0.110     & AGAL338.567+00.109 &  \\
G010.958+0.022 & AGAL010.957+00.022 & G338.850+0.409     & AGAL338.851+00.409 &  \\
G011.034+0.062 & AGAL011.034+00.061 & G339.282+0.136     & AGAL339.283+00.134 &  \\
G012.625-0.017 & AGAL012.623-00.017 & G339.476+0.185     & AGAL339.476+00.184 &  \\
G012.889+0.489 & AGAL012.888+00.489 & G339.582-0.127     & AGAL339.584-00.127 &  \\
G013.179+0.061 & AGAL013.178+00.059 & G339.622-0.121     & AGAL339.623-00.122 &  \\
G014.631-0.577 & AGAL014.632-00.577 & G340.249-0.046     & AGAL340.249-00.046 &  \\
G305.799-0.245 & AGAL305.799-00.244 & G340.785-0.096     & AGAL340.784-00.097 &  \\
G309.384-0.135 & AGAL309.384-00.134 & G341.218-0.212     & AGAL341.217-00.212 &  \\
G311.947+0.142 & AGAL311.947+00.142 & G341.276+0.062     & AGAL341.274+00.061 &  \\
G324.923-0.568 & AGAL324.923-00.569 & G346.480+0.221     & AGAL346.481+00.221 &  \\
G326.608+0.799 & AGAL326.607+00.799 & G346.481+0.132     & AGAL346.481+00.131 &  \\
G326.859-0.677 & AGAL326.859-00.677 & G347.628+0.149     & AGAL347.627+00.149 &  \\
G329.469+0.503 & AGAL329.469+00.502 & G348.884+0.096     & AGAL348.886+00.097 &  \\
G330.283+0.493 & AGAL330.284+00.492 & G349.092+0.105     & AGAL349.091+00.106 &  \\
G331.134+0.156 & AGAL331.134+00.156 & G350.015+0.433     & AGAL350.016+00.432 &  \\
G331.342-0.346 & AGAL331.342-00.347 & G350.520-0.350     & AGAL350.521-00.349 &  \\
G331.442-0.187 & AGAL331.442-00.187 & G350.686-0.491     & AGAL350.687-00.491 &  \\
G331.710+0.603 & AGAL331.709+00.602 & G351.688+0.171     & AGAL351.689+00.172 &  \\
G332.295-0.094 & AGAL332.296-00.094 & G352.604-0.225     & AGAL352.604-00.226 &  \\
G332.364+0.607 & AGAL332.364+00.604 & G352.855-0.201     & AGAL352.856-00.202 &  \\
G332.560-0.148 & AGAL332.559-00.147 & G354.615+0.472     & AGAL354.616+00.472 &  \\
G332.942-0.686 & AGAL332.942-00.686 & G354.724+0.300     & AGAL354.724+00.301 &  \\
G333.163-0.101 & AGAL333.161-00.099 & G355.538-0.105     & AGAL355.538-00.104 &  \\
G333.387+0.032 & AGAL333.386+00.032 & G013.657-0.599$^a$ & AGAL013.658-00.599 &  \\
G336.809+0.119 & AGAL336.808+00.119 & G318.948-0.196$^b$ & AGAL318.948-00.197 &  \\
G336.957-0.225 & AGAL336.958-00.224 & G326.474+0.703$^c$ & AGAL326.474+00.702 &  \\
G336.958-0.977 & AGAL336.958-00.977 & G327.393+0.199$^d$ & AGAL327.393+00.199 &  \\
G337.097-0.929 & AGAL337.098-00.929 & G330.876-0.384$^e$ & AGAL330.876-00.384 &  \\
G337.201+0.114 & AGAL337.201+00.114 & G333.314+0.105$^f$ & AGAL333.314+00.106 &  \\
G337.258-0.101 & AGAL337.258-00.101 & G335.586-0.289$^g$ & AGAL335.586-00.291 &  \\
G337.300-0.874 & AGAL337.301-00.874 & G338.281+0.541$^h$ & AGAL338.281+00.542 &  \\
G337.632-0.079 & AGAL337.632-00.079 & G353.463+0.563$^i$ & AGAL353.464+00.562 & \\\hline
\end{tabular}
\caption{The MMB and ATLASGAL names of the 68 sources studied in this paper. The nine sources considered for pixel-by-pixel study are indicated by alphabets $a$ to $i$. These nine sources are referred to as G13, G318, G326, G327, G330, G333, G335, G338 and G353 respectively}\label{fullsources}
\end{table*}

For lines whose optical depth is known through fitting of hyperfine structure, equation~\eqref{eq:tex} can be used to determine the excitation temperature. The excitation temperature can also be determined for lines that are known to be very optically thick by assuming $\tau$ to be infinity in equation~\eqref{eq:tex}. For the molecules that do not possess hyperfine transitions, we followed the method described in \citet{sanhueza2012chemistry} to estimate the optical depths. The process essentially involves computing the ratio of peak main beam temperatures of an optically thick and optically thin line and then equating it with the ratio of their optical depths. Once we obtain the optical depths, we estimate their excitation temperatures using equation~\eqref{eq:tex}.

In order to study the chemical evolution within the maser host, we performed a pixel-by-pixel SED fitting of molecular data cubes. For this purpose, we chose sources where at least 30 pixels in the moment zero map of MALT90, have S/N ratio~$\ge5$. This results in a sample of 9 sources that are used for what we refer to as the ``pixel-by-pixel analysis''. We also aim to study properties of methanol maser hosts from a chemical perspective compared to sources belonging to different evolutionary stages. Since this cannot be accomplished with just 9 sources, we expanded our source sample to the ones where the brightest pixel in the moment zero map has S/N~$>3$. It is to be noted that the peak signal to noise ratio of individual spectra are much higher since the moment zero map is obtained from the total intensity of the full spectrum of the spectrometer sub-band, with most spectral channels being devoid of any signal. We call this as the statistical analysis of maser sources. 59 sources, apart from the 9 sources listed above, were chosen for the latter approach. In total, our sample constitutes of 68 6.7~GHz methanol maser sources (hereafter MM sources), where we carry out a pixel-by-pixel analysis (approach I) for 9 sources and a statistical analysis (approach II) for the entire source sample. A comparison of the distances and methanol maser luminosities of our sample with that of the 320 sources in Paper~I as well as the 972 sources in the entire MMB catalogue shows our sample to be representative of the methanol maser population in the Galaxy with no systematic biases being introduced by our sample selection procedure (see Appendix~\ref{comp68}). The MMB names and ATLASGAL names of the entire sample of sources are listed in Table~\ref{fullsources}. Among the 68 maser sources, we have 51 N$_{2}$H$^{+}$ detections, 59 HCO$^{+}$, 57 HNC and 52 HCN detections.

\begin{figure}
\begin{center}
\includegraphics[width=0.4\textwidth, trim= 0 0.1cm 0 0]{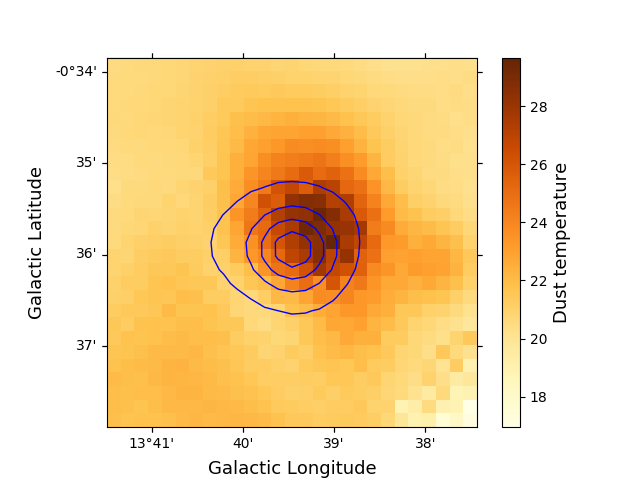}
\includegraphics[width=0.4\textwidth, trim= 0 0.1cm 0 0]{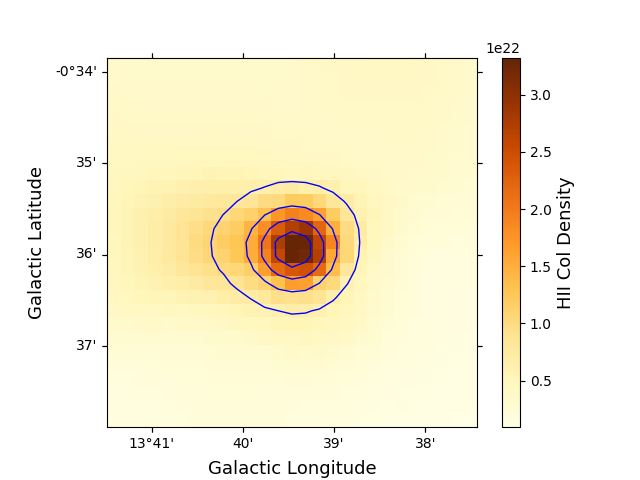}
\caption{An example of temperature(top) and H$_{2}$ column density (bottom) maps obtained by pixel-by-pixel fitting of dust continuum data for the source G13. The SPIRE 250 $\mu$~m emission is overlaid as contours.} \label{fig:tempHmap}
\end{center}
\end{figure}

\subsection{Dust continuum data}

The molecular abundance is calculated by dividing the molecular column density by the H$_2$ column density. The H$_2$ column density is determined using the spectral energy distribution of the dust continuum emission. We made use of the  70, 160~$\mu$m (PACS) as well as 250, 350 and 500~$\mu$m (SPIRE) data, observed as a part of the Herschel infrared Galactic plane survey (Hi-GAL, \citet{molinari2010hi}). The 870~$\mu$m data is obtained from the ATLASGAL survey \citep{schuller2009atlasgal}. The maps at different wave bands have different data units, resolution and plate scales. Moreover, in order to calculate the molecular abundances, it is important to ensure that the H$_2$ column densities are determined at the same world coordinates as that of the MALT90 data. We hence processed the data using the Herschel Interactive Processing Environment (HIPE)\footnote{http://herschel.esac.esa.int/hipe/} using the following steps: Firstly, the surface brightness unit of all the images were converted to Jy pixel$^{-1}$ using the task ‘Convert Image Unit’. The data were then projected onto a common grid, with the pixel size and resolution of the MALT90 data cube. This made sure that each coordinate of the dust continuum map corresponds to the same location in the MALT90 data cube. The plug-in ‘Photometric Convolution’ is used for this purpose. A constant background, estimated from the mean continuum in a region devoid of source emission, was then subtracted from the data. The pixels were then fit using the standard grey body model. We derive the hydrogen column density ($\mathrm{N(H_{2})}$) using the equations discussed in Paper~I. The maps obtained for source G13 is shown in Fig.~\ref{fig:tempHmap} as an example.

For the pixel-by-pixel analysis, we modelled the spectra corresponding to each pixel of the MALT90 data cube and then compared with the dust temperatures and H$_{2}$ column densities of same pixels obtained by pixel-by-pixel SED fitting of dust continuum maps. For the statistical study, we modelled the brightest pixel in the molecular data cube (brightest in the moment zero map) and compared with the H$_{2}$ column density corresponding to that pixel. The dust temperatures for the latter are taken from Paper~I.

\section{RESULTS}\label{sec3}
We present the results obtained from pixel-by-pixel study as well as the statistical study of the sources. We mainly focus on the column densities and abundances of molecules obtained after line modelling. We also briefly account for the integrated line intensities of the molecular emissions. The uncertainties in the column densities and abundances obtained are typically less than 20 percent.
\begin{table*}
\caption{Column densities and abundances.}\label{stat}
         \label{cdens}
  \begin{threeparttable}
  \begin{tabular}{cccccc}
    \hline
    \hline
Quantity & {Min}  &{Max} &{Mean} &{Std.} &{Median} \\
      \hline
N$_I$(HCN) &7.56(12) &9.77(13) &3.39(13) &1.97(13) &2.98(13) \\
X$_I$(HCN) &1.51(-10) &1.32(-8) &1.74(-9) &1.81(-9) &1.29(-9) \\
N$_{II}$(HCN) &1.20(13) &1.06(14) &3.21(13) &2.11(13) &2.66(13) \\
X$_{II}$(HCN) &1.68(-10) &1.06(-8) &2.27(-9) &2.05(-9) &1.56(-9) \\
N$_I$(HNC) &1.59(12) &1.57(14) &2.14(13) &1.95(13) &1.45(13) \\
X$_I$(HNC) &4.49(-11) &2.45(-8) &1.47(-9) &1.70(-9) &1.14(-9) \\
N$_{II}$(HNC) &5.25(12) &8.89(13) &2.12(13) &5.76(12) &1.83(13) \\
X$_{II}$(HNC) &2.24(-10) &4.64(-9) &1.28(-9) &8.27(-10) &9.47(-10) \\
N$_I$(HCO$^+$) &2.72(12) &9.69(13) &1.83(13) &1.81(13) &1.20(13) \\
X$_I$(HCO$^+$) &1.76(-10) &1.53(-8) &1.37(-9) &1.81(13) &1.06(-9) \\
N$_{II}$(HCO$^+$) &4.36(12) &4.82(13) &1.80(13) &1.16(13) &1.42(13) \\
X$_{II}$(HCO$^+$) &1.35(-10) &4.79(-9) &1.26(-9) &1.00(-9) &9.61(-10) \\
N$_I$(N$_2$H$^+$) &5.29(12) &9.25(13) &2.97(13) &1.72(13) &2.50(13) \\
X$_I$(N$_2$H$^+$) &1.33(-10) &7.84(-9) &1.85(-9) &1.41(-9) &1.42(-9) \\
N$_{II}$(N$_2$H$^+$) &2.31(11) &6.84(13) &2.97(13) &1.51(13) &2.48(13) \\
X$_{II}$(N$_2$H$^+$) &1.58(-11) &4.09(-9) &1.81(-9) &1.01(-9) &1.53(-9) \\
    \hline
  \end{tabular}
\begin{tablenotes}
  \item[] a(b) indicates a$\times$10$^{b}$. N and X represents column density and abundance respectively. Subscripts I and II indicates pixel-by-pixel study and statistical study.
\end{tablenotes}
\end{threeparttable} 
\end{table*}

\begin{table}

\caption{Spearman correlation coefficients.}
  \begin{center}

         \label{Spearman}
  \begin{threeparttable}
  \begin{tabular}{ccc}
      
    \hline
    \hline

Molecules & {Pix-by-pix}  &{Stat}  \\
      \hline
HCN &-0.62 &-0.64 \\
HNC &-0.53 &-0.54\\
HCO$^+$ &-0.61 &-0.62 \\
N$_2$H$^+$ &-0.61 &-0.55 \\
\hline
  \end{tabular}
\begin{tablenotes}
  \item[] Correlation coefficients between different molecular species and H$_2$ column densities for pixel-by-pixel study and statistical study.
\end{tablenotes}

\end{threeparttable} 
  \end{center}

\end{table}

\subsection{Dust temperatures and H$_2$ column densities}

The dust temperatures obtained after pixel-by-pixel SED fitting ranges from 13~K to 49~K with a mean value 24~K. The H$_2$ column densities derived for each pixel lie in the range, $(0.16-10.4)\times10^{22}$ cm$^{-2}$ with an average value of $2.68\times10^{22}$ cm$^{-2}$. For the statistical analysis, the dust temperature (taken from Paper I) ranges between 13~K and 36~K with a mean value of 23~K and the H$_2$ column densities lie in the range $(0.25-6.98)\times10^{22}$ cm$^{-2}$ with an average value of $1.86\times10^{22}$ cm$^{-2}$. These values are comparable to what are typically seen in massive star forming regions. For example, the dust temperature and H$_2$ column density values derived by \citet{hoq2013chemical} for 332 high mass clumps are 6.7$-$41.5~K (22~K on average) and $(0.2-93.99)\times10^{22}$ cm$^{-2}$ $(4.89\times10^{22}$ cm$^{-2}$ on average) respectively. The dust temperature values are also consistent with what have been reported by \citet{guzman2015far} for MALT90 clumps (See Appendix~\ref{temp68}).

\subsection{Optical depths and excitation temperatures}
The fits to the hyperfine structure of HCN (1$-$0) show the line to be optically thick with typical values of $\tau$ being greater than 5. In contrast, N$_2$H$^+$ (1$-$0) was found to be typically optically thin, with the optical depth ranging from 0.1 to 1.2 and a mean value of 0.3. We derived the optical depths for HCO$^{+}$ and HNC lines using the optically thin  N$_{2}$H$^{+}$ line, as their optically thin isotopologues (H$^{13}$CO$^+$ and HN$^{13}$C) were undetected in our sources and found both lines to be optically thick. The excitation temperatures calculated for HCN, ranges from 4 to 10~K. We employed the same excitation temperatures for modelling HNC as well, as the HCN and HNC molecules show a tight correlation in their integrated intensity. The two molecules are also observed to have formed via similar chemical channels. The excitation temperatures of HCO$^{+}$ and N$_{2}$H$^{+}$ range from 6$-$12~K and 17$-$30~K respectively.

\subsection{Anomalies in the HCN spectra}
\begin{figure}
\begin{center}
\includegraphics[width=0.44\textwidth, trim= 0.6cm 0.6cm 0.6cm 0.6cm]{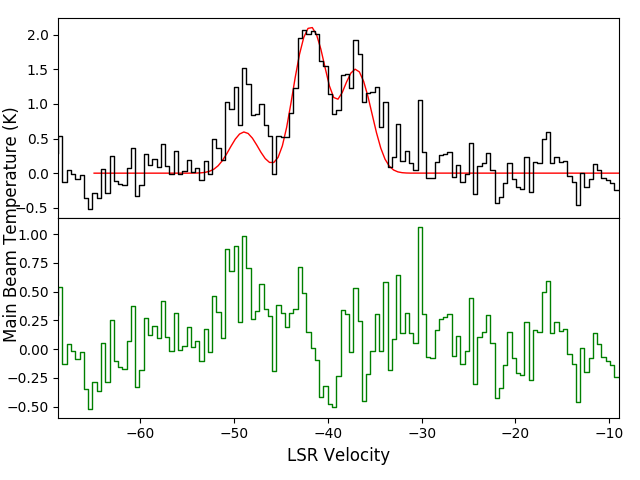}
\caption{An anomalous HCN line profile. The fit obtained (assuming LTE) is shown in red. The fit residual is marked in green.} \label{fig:anomaly}
\end{center}
\end{figure}

Most of the HCN spectra modelled in this work exhibit anomalous and asymmetric line profiles. The complex line profiles are mostly due to hyperfine anomalies observed in the $J=1-0$ line, and also in part due to complex motions within the source. Anomalies in the hyperfine structure of HCN have been observed by several authors to date \citep{guilloteau1981thermal,loughnane2012observations,mullins2016radiative}. The $J=1-0$ transition of HCN has three hyperfine lines ($F=1-0, 1-1$ and $2-1$). Under LTE conditions, the ratios of the relative intensities of the two nearby hyperfine lines of HCN(1-0), are in the form 1:5:3. If the spectrum is anomalous, we observe a boosted $F=0-1$ hyperfine line. This line can also be broader than the rest of the hyperfine components. Fig.~\ref{fig:anomaly} shows an example of an anomalous HCN spectrum taken from our sample. 
While the anomalous spectra could in principle be modelled using non-LTE radiative transfer codes such as MOLFIT \citep{moller2017extended}, such an analysis is beyond the scope of our present work. We have modelled the HCN spectra by fitting the 
$F=1-1$ and $2-1$ components under assumption of LTE. The poor fit to the $F=0-1$ hyperfine line is likely to lead to systematic uncertainties in the derived column densities. In order to account for this, we calculated the ratios of area under the hyperfine satellites of the spectrum and the fit. In almost all the cases, the area under the CASSIS fit is $\sim 20\%$ less than area under the actual spectrum. Hence, we infer that the column densities obtained for HCN are underestimated by $20\%$.

\begin{figure*}
\centering
\includegraphics[width=0.4\textwidth, trim= 0 0.4cm 0 0]{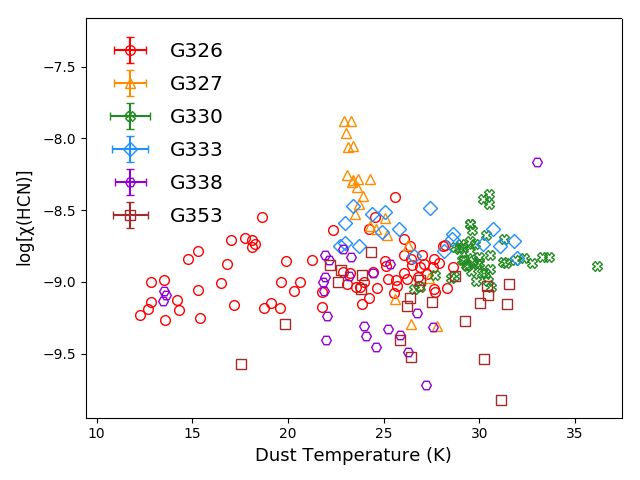}
\includegraphics[width=0.4\textwidth, trim= 0 0.4cm 0 0]{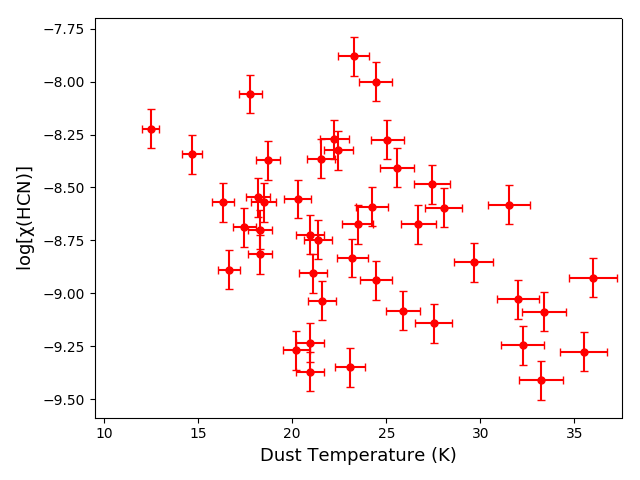}
\includegraphics[width=0.4\textwidth, trim= 0 0.4cm 0 0]{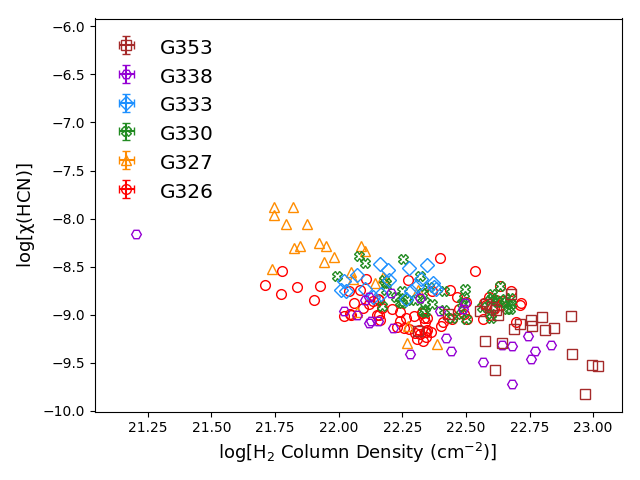}
\includegraphics[width=0.4\textwidth, trim= 0 0.4cm 0 0]{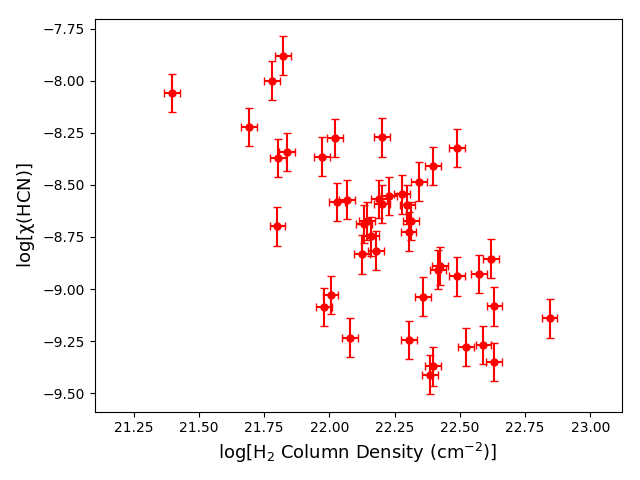}
\includegraphics[width=0.4\textwidth, trim= 0 0.4cm 0 0]{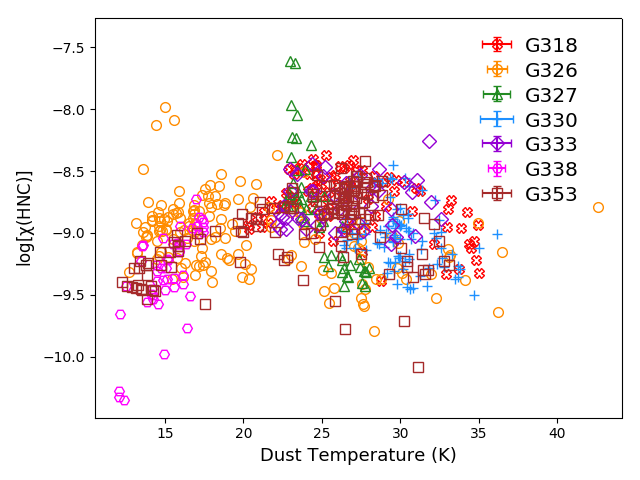}
\includegraphics[width=0.4\textwidth, trim= 0 0.4cm 0 0]{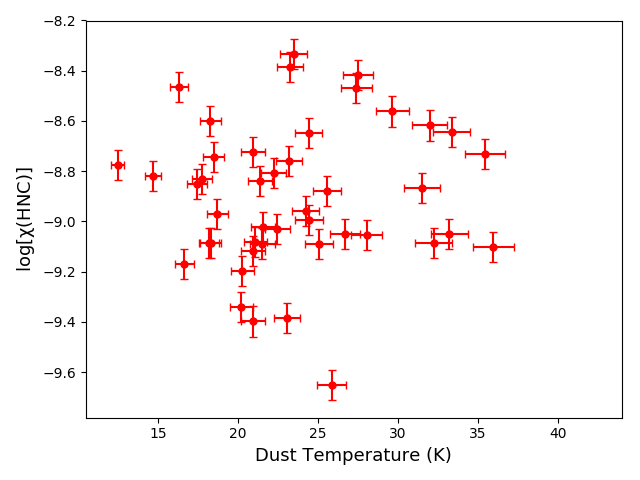}
\includegraphics[width=0.4\textwidth, trim= 0 0.4cm 0 0]{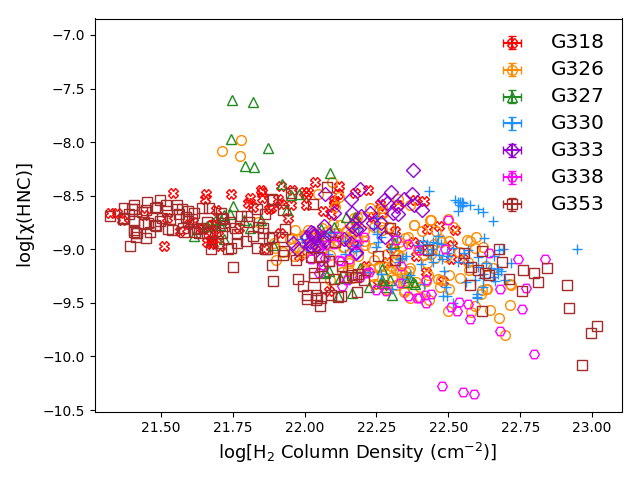}
\includegraphics[width=0.4\textwidth, trim= 0 0.4cm 0 0]{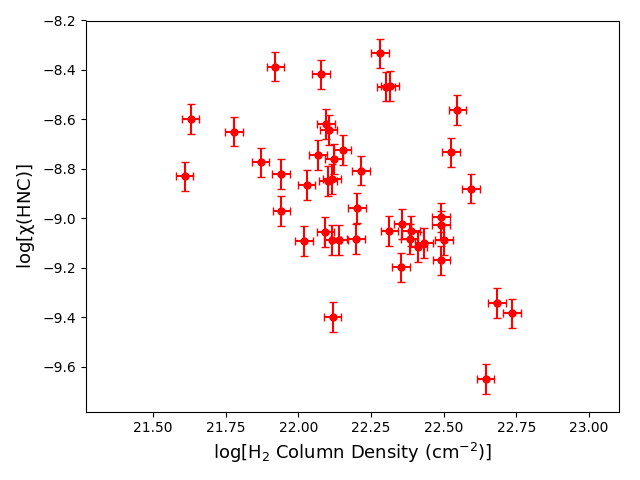}
\caption{Variation of HCN abundance ($X$(HCN)) with respect to dust temperature and H$_{2}$ column density. The first and third panel show the plots for pixel-by-pixel study whereas the second and fourth panel depict the results for statistical study. For the pixel-by-pixel case, only a characteristic error bar is shown in its top corner for the sake of clarity.} \label{fig:HCN}
\end{figure*}

\begin{figure*}

\includegraphics[width=0.4\textwidth, trim= 0 0.4cm 0 0]{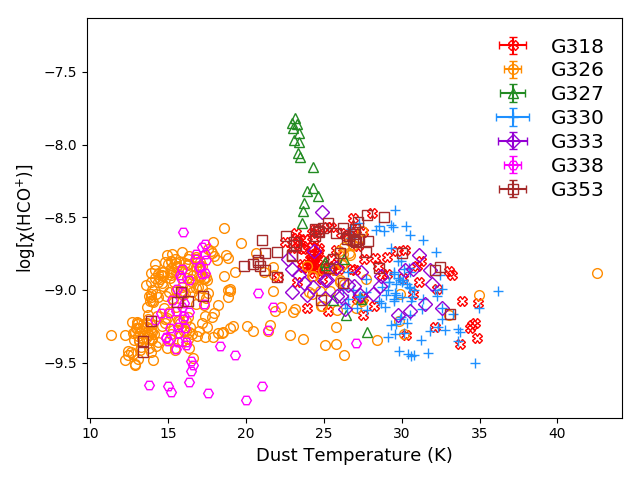}
\includegraphics[width=0.4\textwidth, trim= 0 0.4cm 0 0]{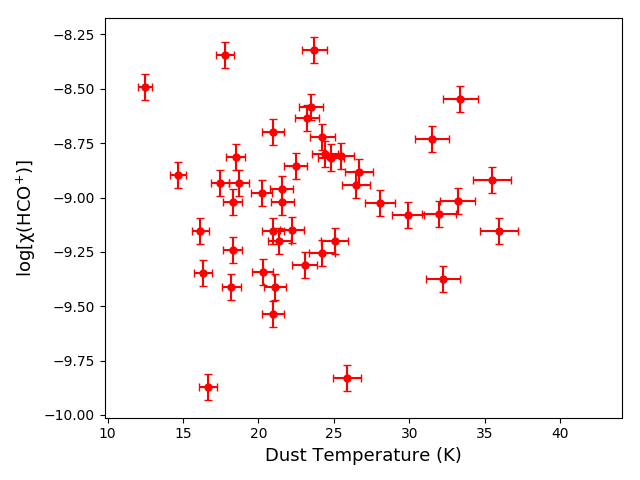}
\includegraphics[width=0.4\textwidth, trim= 0 0.4cm 0 0]{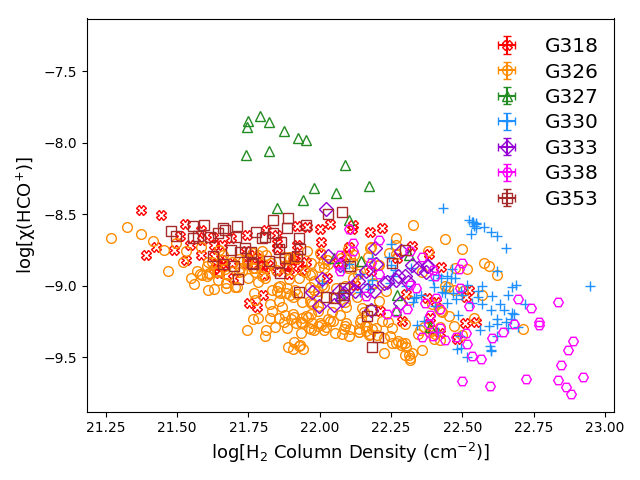}
\includegraphics[width=0.4\textwidth, trim= 0 0.4cm 0 0]{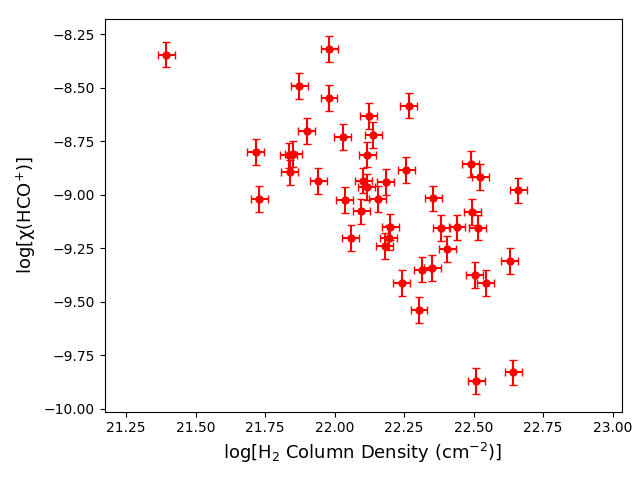}
\includegraphics[width=0.4\textwidth, trim= 0 0.4cm 0 0]{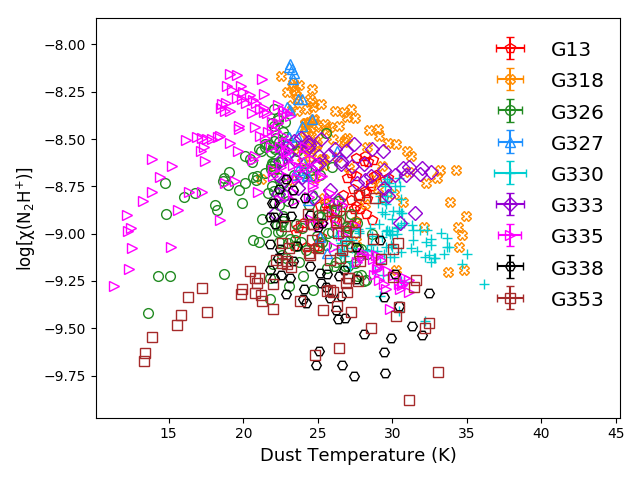}
\includegraphics[width=0.4\textwidth, trim= 0 0.4cm 0 0]{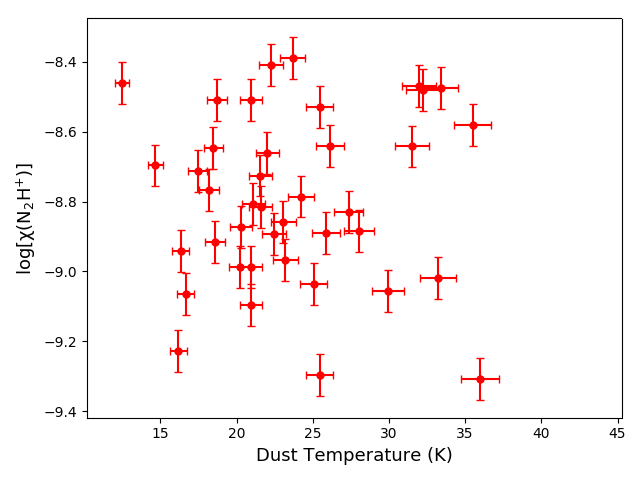}
\includegraphics[width=0.4\textwidth, trim= 0 0.4cm 0 0]{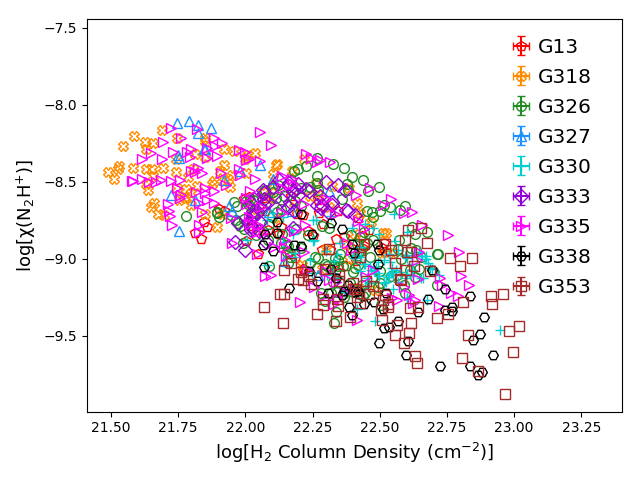}
\includegraphics[width=0.4\textwidth, trim= 0 0.4cm 0 0]{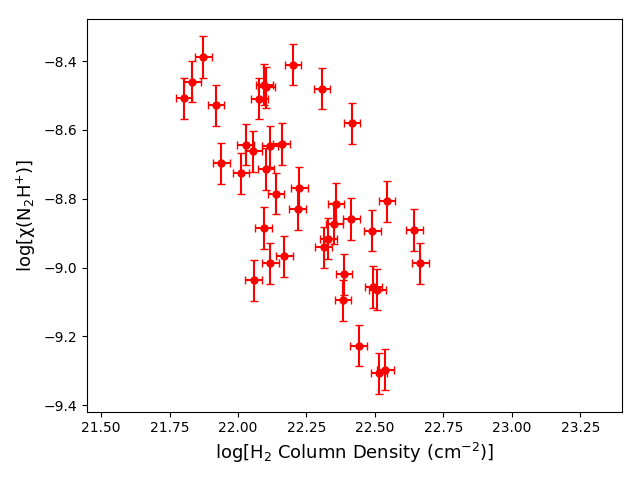}
\caption{Variation of HNC abundance ($X$(HNC)) with respect to dust temperature and H$_{2}$ column density. The first and third panel show the plots for pixel-by-pixel study whereas the second and fourth panel depict the results for statistical study. For the pixel-by-pixel case, only a characteristic error bar is shown in its top-right corner for the sake of clarity.} \label{fig:HNC}
\end{figure*}

\subsection{Molecular column densities and abundances}
Table~\ref{stat} shows the minimum, maximum, mean and median values of molecular column densities and abundances for both the pixel-by-pixel and statistical methods (indicated by I and II respectively). We can see that the column densities and abundances obtained from two methods are very similar to each other. \citet{miettinen2014malt90} reports a similar range for HCN, HNC, HCO$^+$ and N$_2$H$^+$ column densities and abundances for their sample of 35 massive clumps, though their mean HCN and N$_2$H$^+$ column density is an order of magnitude higher than our values.
While the N$_{2}$H$^{+}$ column densities and abundances for all the 51 sources fall in a range similar to that reported by \citet{hoq2013chemical} and \citet{saral2018malt90}, the HCO$^+$ column densities show some deviations.
The HCO$^+$ column densities derived by \citet{hoq2013chemical} for their sample of 333 massive clumps, $(0.4-35.8)\times10^{14}$ cm$^{-2}$ ($3.57\times10^{14}$ cm$^{-2}$ on average), are an order of magnitude higher than those derived for our sources. On the other hand, the median value of HCO$^+$ abundance of our sample ($1.09\times10^{-9}$) is found to be higher than that observed by \citet{zhang2016global} for their sources categorised as $\HII$ regions/PDRs ($3\times10^{-10}$). The mean HCO$^+$ column density value is also less than those derived by \citet{saral2018malt90} for their sample of YSOs and protostellar sources. This could be due to the differences in the methodology adopted for calculating the excitation temperature. \citet{zhang2016global} and \citet{hoq2013chemical} assume the excitation temperature to be equal to the dust temperature. \citet{saral2018malt90} on the other hand, derive excitation temperatures using the optical depth of N$_2$H$^+$ and use the same for the optically thick lines: HCO$^+$, HCN and HNC. In contrast, we find the excitation temperature of HCO$^+$, HCN and HNC to be significantly lower compared to that of N$_2$H$^+$, and much less than the dust temperature.

Figures~\ref{fig:HCN} and \ref{fig:HNC} show HCN, HNC, HCO$^+$ and N$_{2}$H$^{+}$ abundances plotted as a function of dust temperature and H$_2$ column density. While there does not appear to be any trend of the molecular abundances with dust temperature, they appear to be negatively correlated with H$_2$ column density. The correlation coefficients obtained from the Spearman correlation test performed on molecular abundances, and H$_{2}$ column densities, are given in Table~\ref{Spearman}.  We note that the trends shown by the abundances in the statistical study are also reflected in the pixel wise analysis. We also carried out Pearson correlation test on the sample of 68 MM sources for HCN and HNC abundances (with respect to H$_2$ column density) and the coefficients obtained are -0.52 and -0.35 respectively. The Pearson correlation coefficient found between HCN abundance and H$_2$ column density is very similar to that reported by \citet{miettinen2014malt90}, who found the correlation coefficient to be -0.55 for their sample. While \citet{hoq2013chemical} and \citet{miettinen2014malt90} infer that HCO$^+$ abundances tend to increase with evolution, we do not see any such trend in our data. It is to be noted that the inference of HCO$^+$ abundance increasing with evolutionary stage was made by \citet{miettinen2014malt90} assuming that the H$_2$ column density increases with source evolution. This is, however, unlikely to be true since the observation of high H$_2$ column density in single dish studies is more likely to be due to factors other than evolution, such as a larger physical source area being covered in a telescope beam for a more distant source. In contrast, the dust temperature is seen to be a much better probe of the evolutionary stage of the source, with temperatures seen in infrared dark clouds seen to be systematically lower than that in more evolved sources such as methanol masers and $\HII$ regions \citep{giannetti2013physical}. The lack of any correlation between the dust temperature and HCO$^+$ abundance thus casts doubt on the interpretation of \citet{miettinen2014malt90} on the increase of HCO$^+$ abundance with source evolution.

\subsection{Integrated line intensities}

\begin{figure}

\includegraphics[width=0.45\textwidth, trim= 0 0.4cm 0 0]{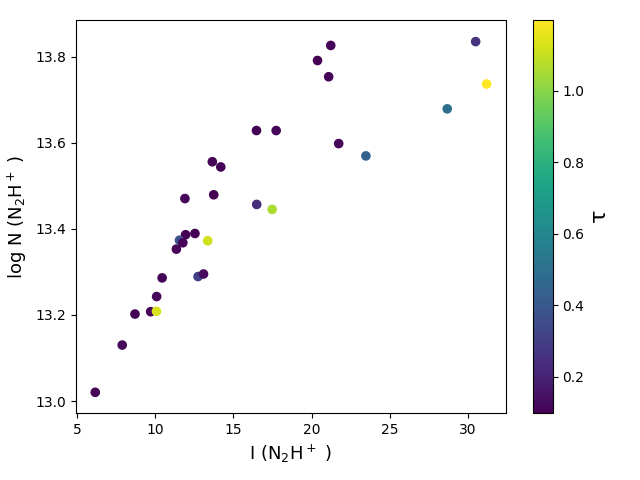}
\includegraphics[width=0.45\textwidth, trim= 0 0.4cm 0 0]{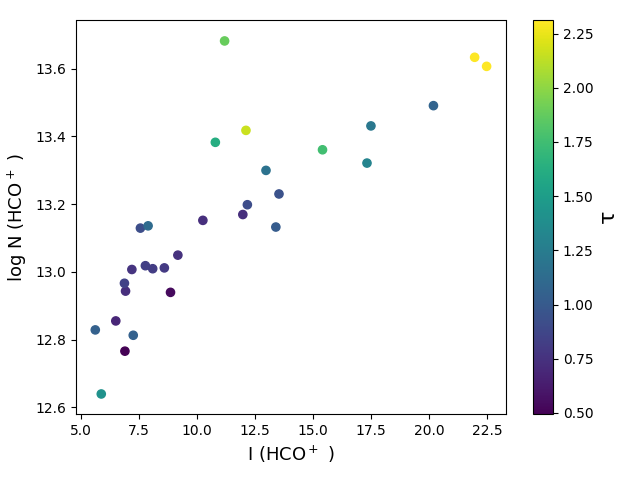}
\caption{Column densities plotted against the integrated intensities of different molecules for MM sources. The colours indicate the variation in optical depth.} \label{fig:moltau}
\end{figure}
Earlier studies \citep[e.g.,][]{rathborne2016molecular,urquhart2019atlasgal} have studied the chemistry in high-mass clumps using integrated line intensities and their ratios. The integrated line intensities divided by H$_2$ column density have also been considered as a rough estimate of molecular abundance \citep{urquhart2019atlasgal}. However, this assumes that the physical conditions are similar for various transitions. In practice, different transitions have different excitation temperatures and optical depths, which result in a variation of the column density from a simple linear relation to the integrated intensity. To illustrate this, Fig.~\ref{fig:moltau} shows the plot of integrated line intensity as a function of column density for N$_2$H$^+$ and HCO$^+$ molecules corresponding to MM sources in our study. The data points are colour-coded by the variation in optical depth. Fig.~\ref{fig:moltau} shows that the sources exhibit a clear bifurcation between high and low optical depth values. The sources possessing lower $\tau$ values tend to show a stronger linear correlation with scatter among these points arising from variations in the excitation temperature. On the other hand, sources with larger optical depth tend to show more dispersion. This implies that integrated intensities are good proxies for column densities only for complex molecules with low optical depth, and studies that target optically thick molecules such as HCN, HNC and HCO$^+$ require radiative transfer modelling in order to obtain reliable column densities. 

The HCN integrated intensities do not show any variation with the dust temperature. On the contrary, the integrated intensity of HNC, shows a weak positive correlation (Spearman correlation coefficient, $r =0.28$). This trend in HNC is in agreement with what is reported by \citet{zhang2016global} and \citet{urquhart2019atlasgal}. The HNC line intensities divided by H$_{2}$ column densities, on the other hand, show no discernible trend. The HCO$^+$ line intensity displays a slight increasing trend with dust temperature ($r = 0.32$), much in agreement with \citet{zhang2016global} and \citet{urquhart2019atlasgal}. The integrated intensity of HCO$^+$ divided by H$_2$ column density show no obvious trends with dust temperature. This is contrary to what is inferred by \citet{urquhart2019atlasgal}, where they detect a weak positive correlation with dust temperature. The integrated intensity of N$_{2}$H$^{+}$ appears to be invariant with dust temperature. This result is in congruence with what is observed by \citet{urquhart2019atlasgal}. \citet{zhang2016global} however, reports a slightly positive correlation. The ratio of integrated line intensity of N$_{2}$H$^{+}$ to the H$_2$ column density show no distinct trend with dust temperature, which is in good agreement with the results of \citet{urquhart2019atlasgal}.

\section{Discussion}\label{sec4}
In this section we discuss in detail the chemistry of different molecules based on the chemical models, followed by a brief discussion on the infall signatures observed in our sample. We also discuss the evolutionary status of methanol maser sources, gleaned from their molecular abundance and line integrated intensity ratios.

\subsection{Comparison to chemical models}
\begin{figure}
\begin{center}
\includegraphics[width=0.4\textwidth, trim= 0 0.4cm 0 0]{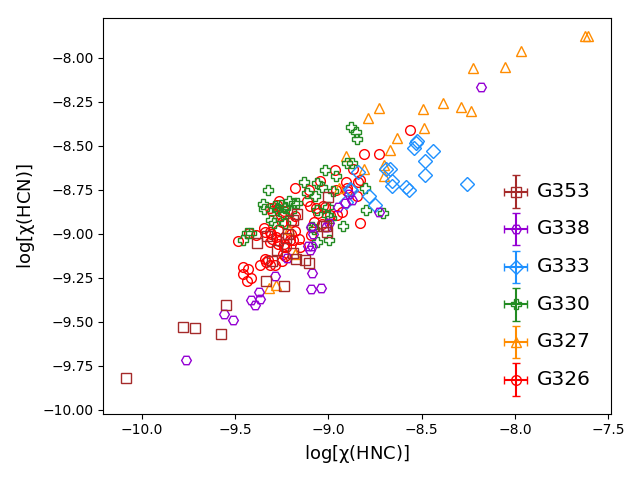}
\includegraphics[width=0.4\textwidth, trim= 0 0.4cm 0 0]{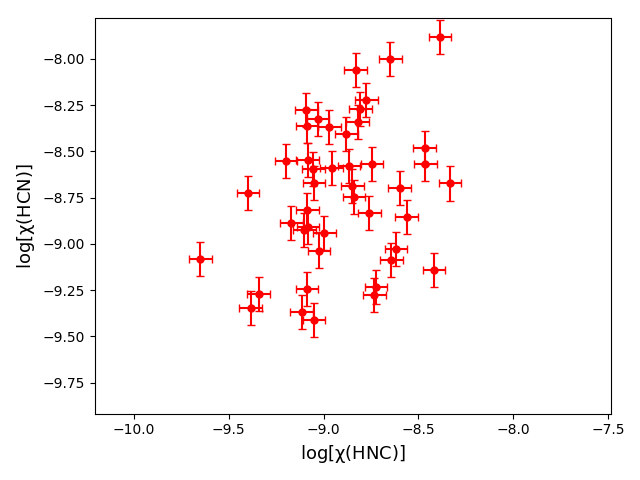}
\caption{$X$(HNC) Vs $X$(HCN). The top panel shows the plots for pixel-by-pixel study whereas the bottom panel depicts the results for statistical study. Pearson correlation coefficient for these two molecular species are 0.72 and 0.35, for pixel-by-pixel and statistical studies respectively.} \label{fig:HCNHNC}
\end{center}
\end{figure}
\begin{figure}
\begin{center}
\includegraphics[width=0.375\textwidth, trim= 0 0cm -0.4cm 0]{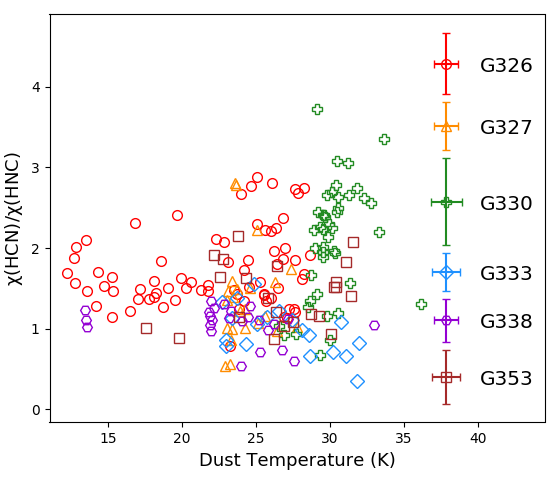}
\includegraphics[width=0.375\textwidth, trim= 0cm 0.4cm 0 0]{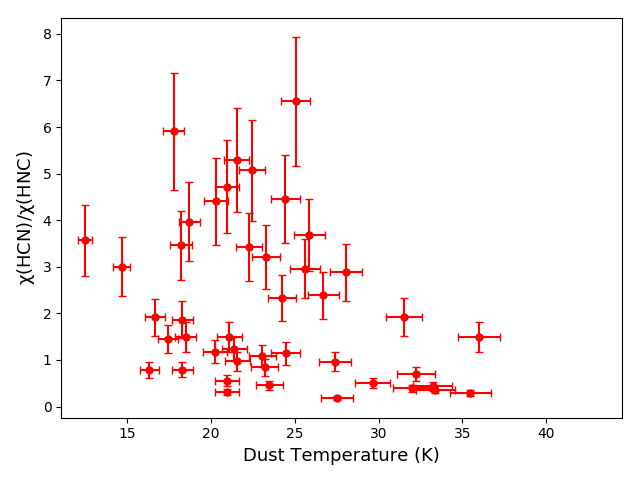}
\caption{$X$(HNC)/$X$(HCN) vs dust temperature. The top panel (a) shows the plots for pixel-by-pixel study whereas the bottom panel (b) shows the results for statistical study.} \label{fig:HCNHNCrat}
\end{center}
\end{figure}

\begin{figure}
\begin{center}
\includegraphics[width=0.47\textwidth, trim= 0 0.4cm 0 0]{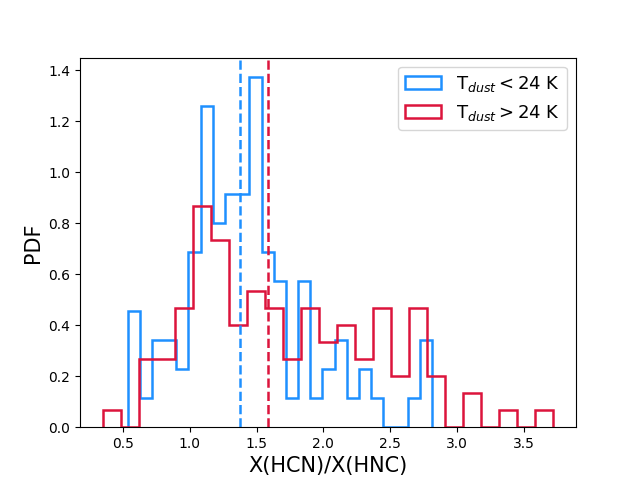}
\caption{Probability density functions of HCN/HNC abundance ratios for $T_{dust}>24$~K and $T_{dust}<24$~K (marked in red and blue respectively). The median values are represented by dashed lines.} \label{fig:pdf}
\end{center}
\end{figure}
\begin{figure*}
 \begin{center}
    \includegraphics[width=0.33\textwidth, trim= 0.2cm 0.4cm 0.2cm 0cm]{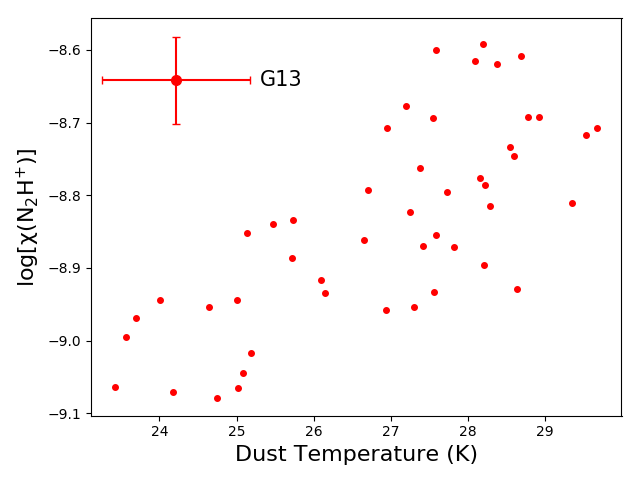}
    \includegraphics[width=0.33\textwidth, trim= 0.2cm 0.4cm 0.2cm 0cm]{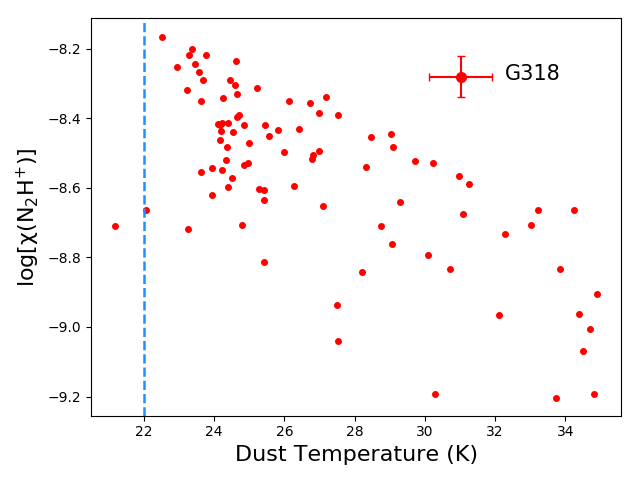}
    \includegraphics[width=0.33\textwidth, trim= 0.2cm 0.4cm 0.2cm 0cm]{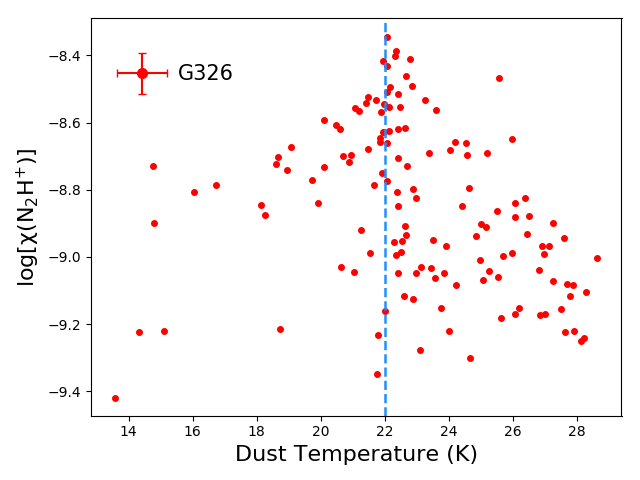}
    \includegraphics[width=0.33\textwidth, trim=0.2cm 0.4cm 0.2cm 0cm]{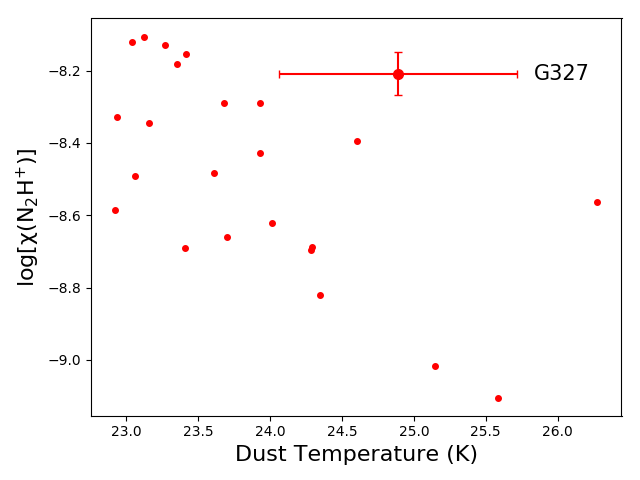}
    \includegraphics[width=0.33\textwidth, trim= 0.2cm 0.4cm 0.2cm 0cm]{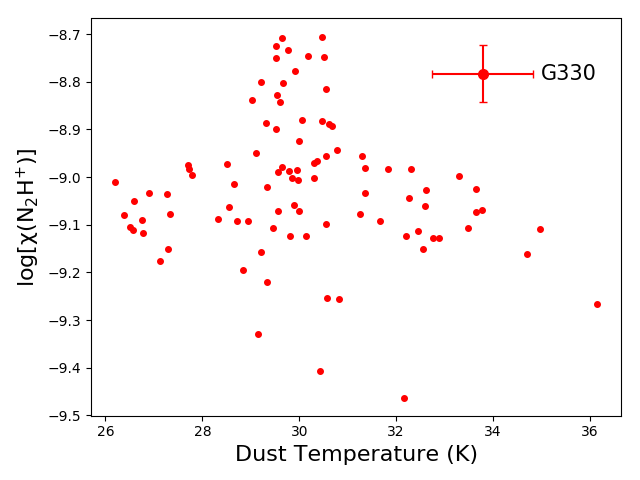}
    \includegraphics[width=0.33\textwidth, trim= 0.2cm 0.4cm 0.2cm 0cm]{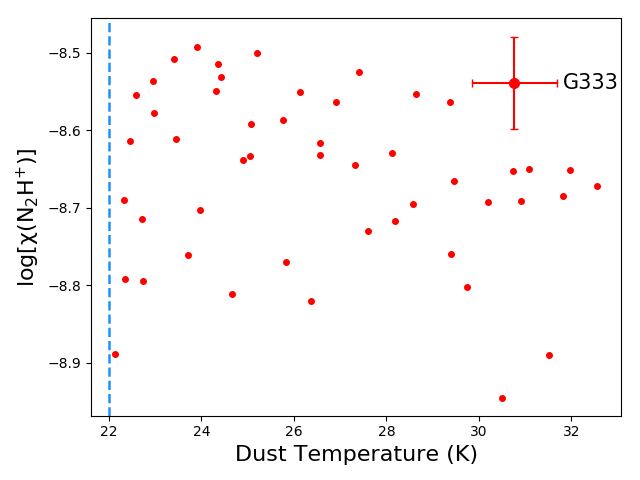}
    \includegraphics[width=0.33\textwidth, trim= 0.2cm 0.4cm 0.2cm 0cm]{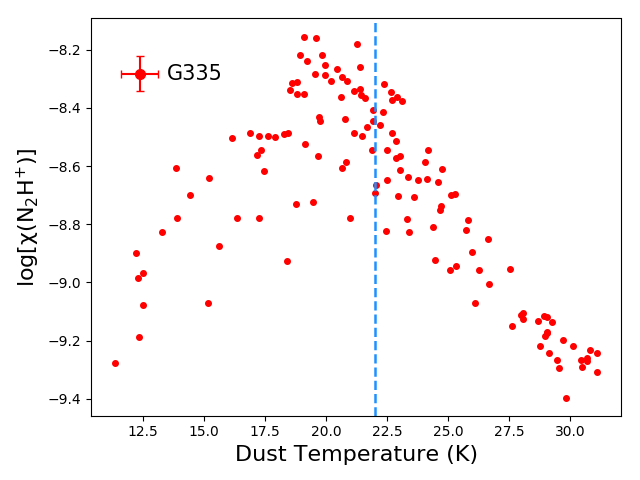}
    \includegraphics[width=0.33\textwidth, trim= 0.2cm 0.4cm 0.2cm 0cm]{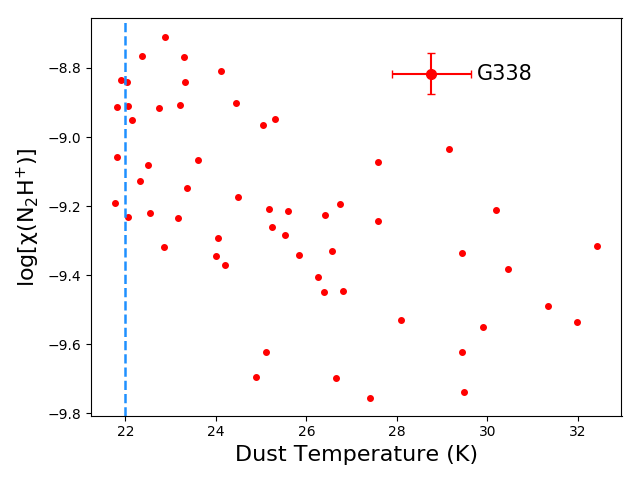}
    \includegraphics[width=0.33\textwidth, trim= 0.2cm 0.4cm 0.2cm 0cm]{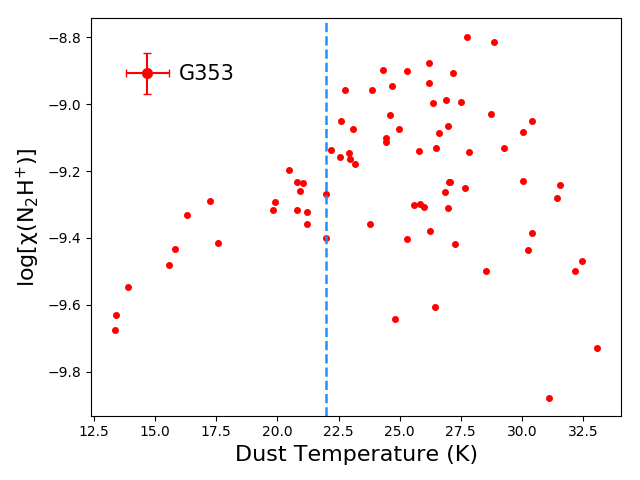}
    \caption{The abundance of N$_2$H$^+$ as a function of dust temperature. The vertical dashed blue line is at T$_{dust}$ = 22 K. Characteristic error bar is shown in the top corner of the plot.} \label{fig:N2H+pix}
 \end{center}
\end{figure*}
\begin{figure}
    \includegraphics[width=0.45\textwidth, trim= 0cm 0cm 0 0cm]{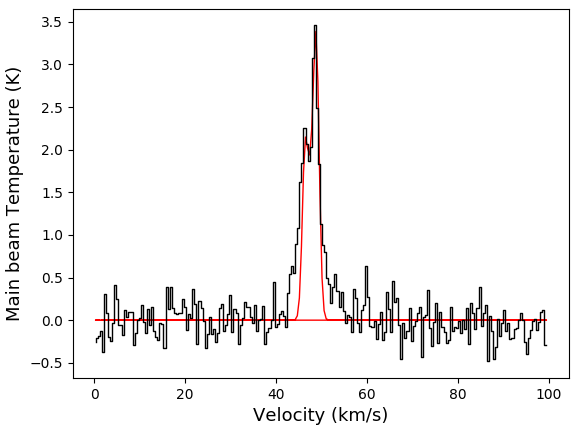}
    \includegraphics[width=0.45\textwidth, trim= 0cm 0cm 0 0cm]{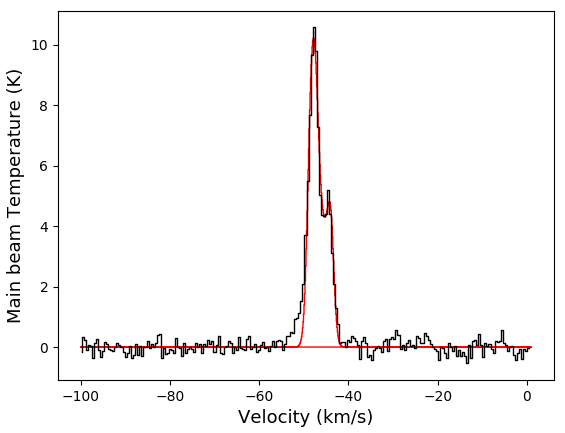}

    \caption{HCO$^+$ spectrum of a single pixel fitted by "Hill5" model for G13 and G335 (top and bottom panels respectively).}
    \label{fig:selfab}
\end{figure}

\begin{figure}
  \includegraphics[width=0.45\textwidth, trim= 0 0.0cm 0 0]{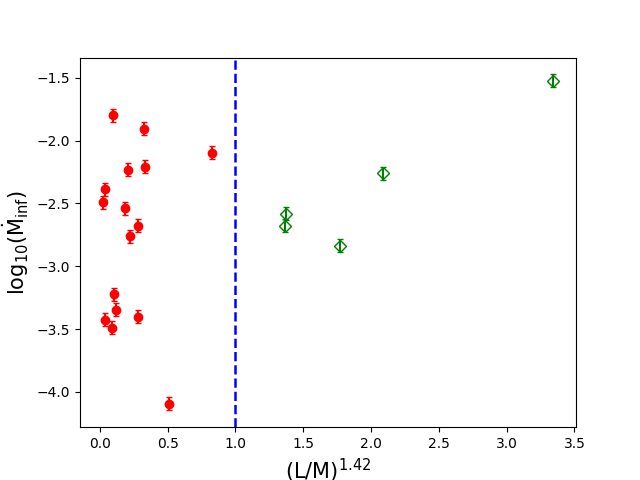}
    \caption{Mass infall rates plotted against (L/M)$^{1.42}$. The methanol maser hosts that are in accretion phase and clearing phase are shown in red (filled circles) and green (unfilled diamonds) points, respectively. The blue dashed line represents the fit to the ‘IR-P’ sources in \citet{molinari2008evolution}.} \label{fig:ml}
\end{figure}
According to gas-phase chemical models, HCN and its metastable geometrical isomer HNC are primarily produced through the  dissociative recombination reaction of HCNH$^+$ as described in Appendix~\ref{chemistryHCN}. Fig.~\ref{fig:HCNHNC} shows HCN abundance plotted as a function of HNC abundance. We can clearly see a strong positive correlation in the pixel-by-pixel case whereas there is more scatter in the statistical study. A Pearson correlation test carried out on the HNC and HCN abundances for the pixel-by-pixel analysis yields a correlation coefficient of 0.72 with 99.8\% confidence level ($p$-value of 0.002)\footnote{Throughout the paper we have adopted a threshold of $>3\sigma$ confidence for claiming statistical significance, which corresponds to a $p$-value $<0.003$ for the null hypothesis of the two samples not being correlated.} For the statistical analysis, we obtained the Pearson correlation coefficient to be 0.35 ($p$-value of 0.0003).
A similar positive correlation is observed by \citet{miettinen2014malt90} (Pearson correlation coefficient being 0.45) for their statistical study of IRDCs. 
The abundance ratio between HCN and HNC is found to lie in the range 0.21 to 4.76 with a mean value of 1.64 (median is 1.49). The mean value is closer to unity as suggested by the chemical models. \citet{miettinen2014malt90} reports their HCN/HNC ratio to be in the range 0.15 to 16.6, with a median value of 1.40. These values agree well with our results. The mean value of abundance ratio determined by \citet{zhang2016global} for their sample of high mass stars is seen to be greater than unity (3.91). However, the HCN/HNC abundance ratio is very sensitive to the adopted dust temperatures \citep{{goldsmith1986molecular},{1992A&A...256..595S}}. While \citet{zhang2016global} have assumed the excitation temperature to be equal to the dust temperature, we find the excitation temperature to be much lower than the dust temperature. The very different abundance ratio of \citet{zhang2016global} may be due to this difference in the value of assumed excitation temperature.

This being said, the HCN/HNC abundance ratio has been shown to vary between regions \citep{1992A&A...256..595S} and this has been attributed to conversion of HNC to HCN at temperatures above $T\sim 24$~K [see equation~\eqref{eq:HNCconsum}]. The abundance ratio appears to show a variation around 24~K for the pixel-by-pixel case (top panel, Fig.~\ref{fig:HCNHNCrat}).  We investigate this further in Fig.~\ref{fig:pdf} which shows the probability distribution functions (PDFs) for the ratios of samples with dust temperatures above and below $T_{dust}=24$~K shown in red and blue, respectively. The sample with $T_{dust}>24$~K has a higher median ratio, which is in agreement with \citet{1992A&A...256..595S} who interpret this increase with source evolution. The lower panel of Fig.~\ref{fig:HCNHNCrat}
shows a decreasing trend with increasing dust temperature, although at $<3\sigma$ confidence level based on a Pearson test.  We would therefore like to caution against over interpretation of this trend.

We also investigate the variation in N$_2$H$^+$ abundance with respect to dust temperature, for our sample of MM sources. Fig.~\ref{fig:N2H+pix} shows the N$_2$H$^+$ abundance plotted as a function of dust temperature in each pixel of the 9 sources (S/N~$\gtrsim 5$). It is seen that for all the sources except G13, the abundances increase till $T_{dust}\simeq22$~K and then decrease as the temperature increases. The increase in N$_2$H$^+$ abundances at lower temperatures ($T<24$~K) is also observed by \citet{sanhueza2012chemistry}, \citet{hoq2013chemical}, \citet{miettinen2014malt90} and several other studies. This positive correlation is often associated with the enhanced N$_2$ evaporation from dust grains, resulting in the formation of more N$_2$H$^+$ molecules \citep{chen2013newly}. As the temperature exceeds 22~K, we see a decline in the abundance of N$_2$H$^+$. This is in agreement with the chemical models, which suggest the liberation of CO molecules from the dust grains at T$_{dust}$ > 20~K, causing the depletion of N$_2$H$^+$ molecules. 

\subsection{Detection of infall signatures}

HCO$^+$ is one of the suitable candidates for investigating infall motions in both low mass and high mass star forming regions. To distinguish self-absorbed line from other double peaked line profiles, we generally make use of its optically thin isotopologue H$^{13}$CO$^+$ \citep{klaassen2012looking}. \citet{mardones1997search} defined a dimensionless asymmetry parameter $\delta$V, to quantify the blue-skewed profiles,
\begin{equation}
   \delta V = \dfrac{ V_{thick}-V_{thin}}{\Delta V}
\end{equation}
Here, $V_{thick}$ and  $V_{thin}$ represent the LSR velocities of the optically thick and optically thin lines and $\Delta V$ is the FWHM of the optically thin line. Line profiles with blue asymmetry are characterized by negative values of the asymmetry parameter -- i.e. $\delta V < 0$. As we do not have any strong (> 3$\sigma$) H$^{13}$CO$^+$ detections, we use optically thin N$_2$H$^+$ for the calculations. We obtained blue-asymmetries for 21 sources in our sample. In addition, two sources, G13 and G335, displayed a blue asymmetric self-absorbed line profile. The line profiles of these sources were modelled using the radiative transfer model "Hill5" \citep{de2005molecular} in order to determine the infall velocity. Fig.~\ref{fig:selfab} illustrates HCO$^+$  fitted by "Hill5" model for a single pixel of G13 and G335. For sources that exhibited blue-asymmetry, but not self-absorption, we used V$_{in}$= $V_{thin}-V_{thick}$ as an estimate of the infall velocity. The mass infall rate is then computed from the infall velocity as
\begin{equation}
    \Dot{M}_{inf} = 3\dfrac{M_{cl}}{R_{cl}} V_{in}
\end{equation}
where $M_{cl}$ and $R_{cl}$ are the mass and radius of the clumps respectively. These values are taken from Table~3 of Paper~I. The mass infall rates calculated, fall in the range $8.0\times10^{-5}$ M$_{\odot}$yr$^{-1}-1.24\times10^{-2}$ M$_{\odot}$yr$^{-1}$ with median value being $2.3 \times10^{-3}$ M$_{\odot}$yr$^{-1}$. The uncertainties in these values are typically less than 15 percent. The calculated infall rates are in good agreement with values that are typically found towards massive star forming regions \citep{young1998radiative,fontani2002structure,lopez2010comparative,saral2018malt90}. Fig.~\ref{fig:ml} shows $\Dot{M}_{inf}$ as a function of luminosity to mass (L/M) ratio. The L/M ratio serves as a diagnostic tool to infer the evolutionary state of the source. In Fig.~\ref{fig:ml}, the x-axis is $(L/M)^{1.42}$, since this represents the fit to the ‘IR-P’ sources in \citet{molinari2008evolution} and delineates sources in accretion phase from those in clearing phase (refer Figure~11 of Paper~I). The sources belonging to the accretion phase and clearing phase are marked in red and green respectively. Although the mass infall rates appear to show a weak positive correlation with L/M ratios ($r=0.3$), this is inconclusive due to $p$-value$>$0.003. A larger sample of sources in clearing phase can elucidate the evolutionary effects of infall rates. \citet{wyrowski2016infall} report a similar result for their sample of high mass clumps, where 3 of their 8 sources showed higher infall rates towards higher L/M. \citet{he2016properties} obtained median mass infall rates of $(7-8)\times10^{-3}$ M$_{\odot}$yr$^{-1}$ for pre-stellar, protostellar and ultra-compact $\HII$ regions for their sample of 732 massive clumps and concluded that the infall rates are independent of the evolutionary stage.  We also found no significant trends in the mass infall rate when plotted against methanol maser luminosities.

\subsection{Implications about the evolutionary stage}

As indicated earlier in $\S1$, a number of studies \citep[e.g.,][and references therein]{ellingsen2006methanol,pandian2010spectral,billington2019atlasgal} using continuum emission from radio to infrared wavelengths strongly suggest that 6.7~GHz methanol masers trace an early evolutionary stage of massive star formation, mostly prior to the formation of a $\HII$ region. In particular, the mass to luminosity ratio of methanol maser sources are consistent with $\sim$93\% of them  being in accretion phase (Paper I). A similar study by \citet{jones2020evolutionary} compares the physical properties of clumps associated with maser emission with that of the protostellar sources without maser emission and find the L/M ratio for maser associated clumps to be slightly higher than that of the latter, suggesting that the maser sources are more evolved. In this context, it is of interest
to see whether the chemical properties of these clumps, examined based on the chemical model, is in agreement with the conclusions derived from their physical properties. Since different molecular species form in specific chemical environments, the ratio of their abundances and intensities can act as a tracer for the evolutionary stage of the high-mass clump. For example, \citet{rathborne2016molecular} and \citet{urquhart2019atlasgal} have observed several line ratios to show systematic trends with source evolution.
\begin{figure}
\begin{center}
\includegraphics[width=0.43\textwidth, trim= 0 0.4cm 0 0]{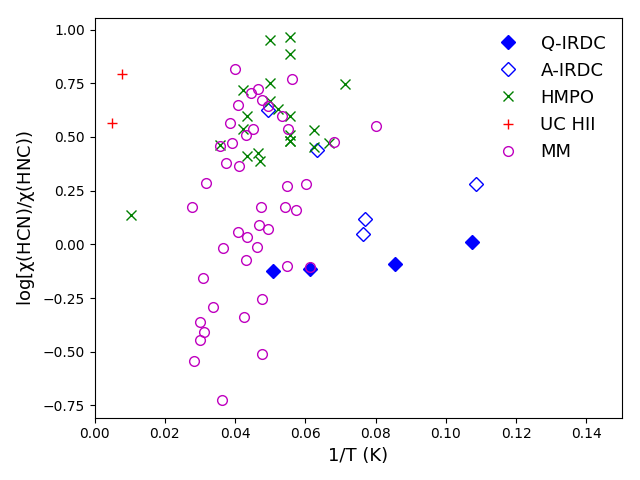}
\includegraphics[width=0.43\textwidth, trim= 0 0.4cm 0 0]{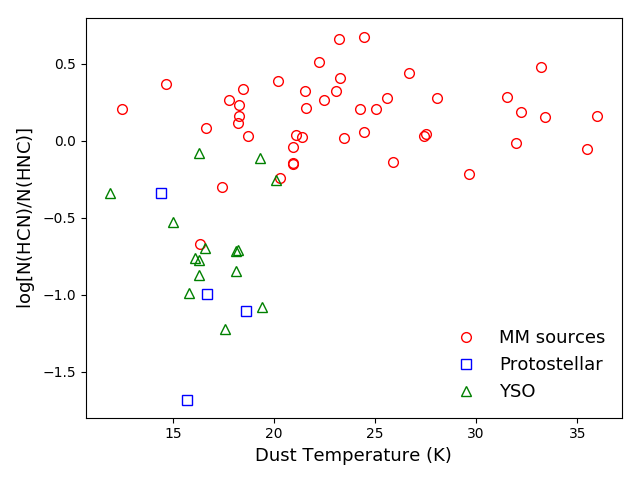}
\caption{MM sources over plotted on results obtained by previous studies. The top panel shows the HCN/HNC ratios derived by \citet{jin2015hcn} and the bottom panel shows that of \citet{saral2018malt90}.}
\label{fig:HCNHNCoverplot}
\end{center}
\end{figure}

\begin{figure}
\begin{center}
\includegraphics[width=0.4\textwidth, trim= 0 0.3cm 0 0]{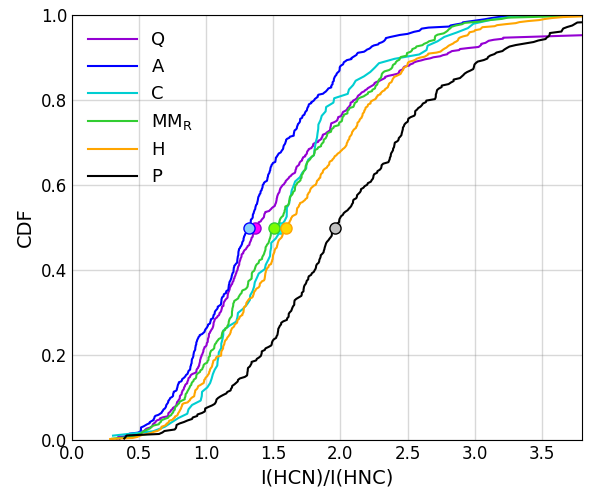}
\includegraphics[width=0.4\textwidth, trim= 0 0.4cm 0 0]{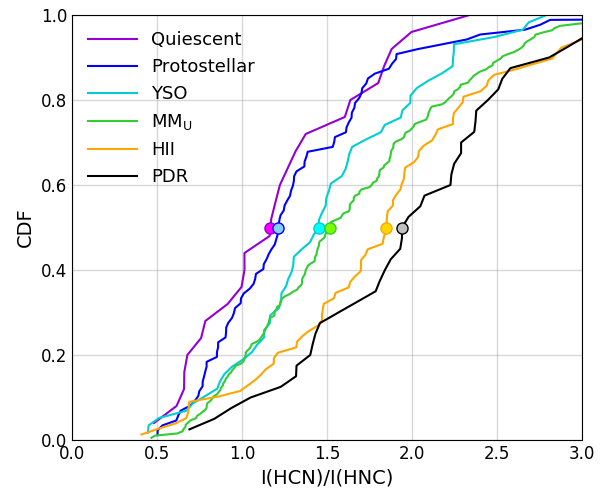}
\caption{MM sources compared with the HCN/HNC integrated intensity ratios of \citet{rathborne2016molecular} and \citet{urquhart2019atlasgal} shown in top and bottom panels respectively. The filled circles on the CDFs represent the median line intensity ratios. Letters ‘Q’, 'A', 'C', 'H' and 'P' denotes 
quiescent, protostellar, compact  $\HII$ regions, extended  $\HII$ regions and  photo-dissociation regions respectively.} \label{fig:HNCHCNcomp}
\end{center}
\end{figure}
We focus on four molecular abundance (intensity) ratios: HCN/HNC, HNC/HCO$^+$, N$_2$H$^+$/HCO$^+$ and N$_2$H$^+$/HNC. Here we compare these abundance ratios of our sources with those that are classified to be in different evolutionary stages by earlier studies. We also compare the line intensity ratios of the methanol maser sources with the overall sample of \citet{rathborne2016molecular} and \citet{urquhart2019atlasgal}. In this regard, it should be noted that although our data is from the MALT90 survey, our integrated line intensities are in general higher than that of \citet{rathborne2016molecular}. This is primarily due to the differences in spectrum extraction from the data cubes. While we have extracted our spectra from the brightest pixel as discussed in Section 2, \citet{rathborne2016molecular} use an averaged spectrum, obtained by averaging over 9 pixels around the dust peak. This has the effect of reducing the intensity of the spectrum although the signal to noise ratio can be higher. Our integrated line intensities are also found to be frequently different from those of \citet{urquhart2019atlasgal} for the same sources, with the discrepancy most likely arising from differences in the pointing location and velocity resolution ($\sim$ 0.9~km s$^{-1}$, which is worse than that of MALT90). We hence do not compare the line ratios determined in our work with that of \citet{rathborne2016molecular} and \citet{urquhart2019atlasgal}. Rather, we extract the molecular line intensities of methanol maser detections in the MMB from \citet{rathborne2016molecular} and \citet{urquhart2019atlasgal} and compare their ratios with that of sources in other evolutionary states. This is done separately for the sample of \citet{rathborne2016molecular} and \citet{urquhart2019atlasgal}. We will refer to the maser sample separated out from \citet{rathborne2016molecular} and \citet{urquhart2019atlasgal} as MM$_R$ and MM$_U$ respectively, for the sake of clarity. The minimum, maximum and median values of column density and abundance ratios between different molecules for 68 MM sources are given in Table~\ref{ratios}.

\begin{table}

\caption{Molecular ratios}
  \begin{center}

       \label{ratios}
  \begin{threeparttable}
  \begin{tabular}{cccc}

   \hline
Molecular ratios & {Min}  &{Max} &{Median}  \\
      \hline
N(HCN)/N(HNC) & 0.21 	 & 4.76	 & 1.49 	 \\
X(HCN)/X(HNC) & 0.21 	 & 4.76	 & 1.50 	 \\
N(HNC)/N(HCO$^+$) & 0.37 	 & 7.68 	 & 1.17 	 \\
X(HNC)/X(HCO$^+$) & 0.29 	 & 7.68 	 & 1.16 	 \\
N(N$_2$H$^+$)/N(HCO$^+$) & 0.03 	 & 8.71 	 & 1.66	  \\
X(N$_2$H$^+$)/X(HCO$^+$) & 0.03 	 & 8.71 	 & 1.58	  \\
N(N$_2$H$^+$)/N(HNC)  & 0.02 	 & 5.73 	 & 1.40 	 \\
X(N$_2$H$^+$)/X(HNC)  & 0.02 	 & 5.73 	 & 1.47 	 \\

\hline
  \end{tabular}
\begin{tablenotes}
  \item[] Minimum, maximum and median values of different molecular ratios for 68 MM sources. N and X represents column density and abundance respectively.
\end{tablenotes}

\end{threeparttable} 
  \end{center}

\end{table}

We begin by comparing the HCN/HNC abundance ratios with the results of \citet{jin2015hcn} and \citet{saral2018malt90}. The former estimate the abundance ratio for Infrared dark clouds (IRDCs, quiscent or active, depending on star formation activity), High-mass protostellar objects (HMPOs) and Ultra-compact $\HII$ regions (UCHIIs). The latter classify their sources into three evolutionary stages: massive star forming regions (MSF), young stellar objects (YSO) and protostellar. The top panel of Fig.~\ref{fig:HCNHNCoverplot} shows that MM sources have similar temperatures as HMPOs but their abundance ratios partially span the IRDC region as well.  
In contrast, the bottom panel of Fig.~\ref{fig:HCNHNCoverplot} shows the abundance ratio to be larger than YSOs and protostellar sources. Fig.~\ref{fig:HCNHNCoverplot} also shows that the abundance ratio inferred by \citet{jin2015hcn} differs systematically from that of sources in similar evolutionary stage (e.g. HMPO vs Protostellar or YSO) in \citet{saral2018malt90}. This may be due to the abundance ratio being determined by comparing the column densities of H$^{13}$CN and HN$^{13}$C by \citet{jin2015hcn}. Considering that the methodology that we have used in our work is similar to that of \citet{saral2018malt90}, the methanol masers appear to be tracing sources that are more evolved than the protostellar phase. However, due to the small sample size, this result is not statistically significant as inferred by a Kolmogorov-Smirnov (KS) test\footnote{The null hypothesis considered for KS tests and Student's T tests in this paper is of the two samples being drawn from the same population. We infer two samples to be disparate only 
if the $p$-value$<$0.003, which corresponds to a $>3\sigma$ confidence level.}. In Fig.~\ref{fig:HNCHCNcomp} we compare the cumulative distribution functions (CDFs) of HCN/HNC line intensity ratios of MM$_R$ and MM$_U$ sources with quiescent, protostellar, $\HII$ regions and photo-dominated regions (PDRs) in \citet{rathborne2016molecular} (top panel) and \citet{urquhart2019atlasgal} (bottom panel), respectively. The median value of the ratio for methanol maser sources is seen to be greater than that of protostellar sources, but smaller than that of $\HII$ regions/PDRs. Since the HCN/HNC line ratio is seen to increase with source evolution, this suggests that the methanol maser sources are on average more evolved than protostellar sources but less evolved than $\HII$ regions/PDRs. The KS test shows the maser sample to be distinct from protostellar sources but not $\HII$ regions, showing the former to be a stasistically significant result.

\begin{figure}
\begin{center}
\includegraphics[width=0.4\textwidth, trim= 0 0.3cm 0 0]{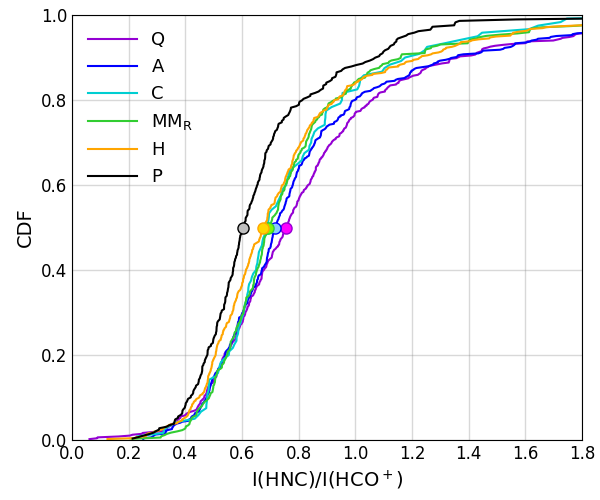}\quad
\includegraphics[width=0.4\textwidth, trim= 0 0.4cm 0 0]{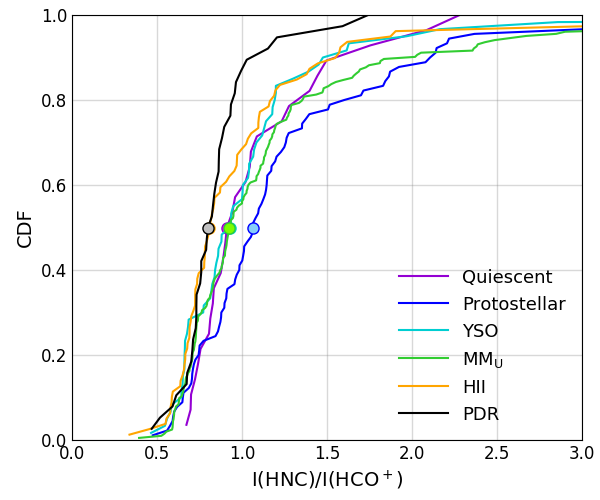}
\caption{Methanol maser sources compared with the HNC/HCO$^+$ integrated intensity ratios of \citet{rathborne2016molecular} and \citet{urquhart2019atlasgal} shown in top and bottom panels respectively. The filled circles on the CDFs represent the median line intensity ratios. Letters ‘Q’, 'A', 'C', 'H' and 'P' denotes 
quiescent, protostellar, compact $\HII$ regions, extended $\HII$ regions and  photo-dissociation regions respectively.} \label{fig:HNCHCOPcomp}
\end{center}
\end{figure}

\citet{miettinen2014malt90} compute HNC/HCO$^+$ column density ratios for IR-dark sources and IR-bright sources, whereas \citet{zhang2016global} obtain HNC/HCO$^+$ abundance ratios of prestellar, protostellar and $\HII$/PDR clumps. Both of these studies imply that the  HNC/HCO$^+$ ratios decrease as the source evolves. On comparing the values for MM sources with these datasets, we find a higher median column density ratio than IR-dark sources and a lower median abundance ratio than $\HII$/PDR sources. 
Fig.~\ref{fig:HNCHCOPcomp} shows the line intensity ratio of HNC/HCO$^+$ for the classifications of \citet{rathborne2016molecular} and \citet{urquhart2019atlasgal}. The line intensity ratio is seen to decline with source evolution, similar to the observation with ratios of column densities. This may be due to enhanced abundance of HNC in colder clumps \citep{hoq2013chemical}. Fig.~\ref{fig:HNCHCOPcomp} shows that the MM$_R$ and MM$_U$ sources have a larger median I(HNC)/I(HCO)$^+$ than both PDRs and $\HII$ regions, and a smaller median ratio than protostellar sources. However, the KS test identifies only methanol maser sources and PDRs to be drawn from seperate populations. Hence, we can ascertain than that the maser sources are less evolved than PDRs. Comparing with Fig.~\ref{fig:HNCHCNcomp}, we can also say that the HNC/HCO$^+$ ratio is not as sensitive as the HCN/HNC ratio in distinguishing between evolutionary states.

\begin{figure}
\begin{center}
\includegraphics[width=0.4\textwidth, trim= 0 0.3cm 0 0]{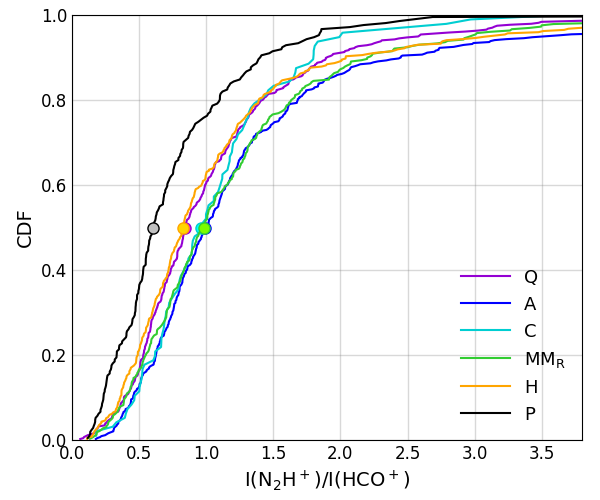}
\includegraphics[width=0.4\textwidth, trim= 0 0.4cm 0 0]{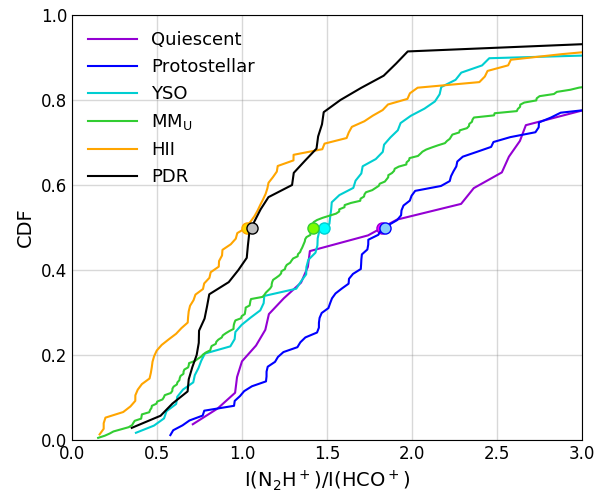}
\caption{MM sources compared with the N$_2$H$^+$/HCO$^+$ integrated intensity ratios of \citet{rathborne2016molecular} and \citet{urquhart2019atlasgal} shown in top and bottom panels respectively. The filled circles on the CDFs represent the median line intensity ratios. Letters ‘Q’, 'A', 'C', 'H' and 'P' denotes 
quiescent, protostellar, compact  $\HII$ regions, extended  $\HII$ regions and  photo-dissociation regions respectively.} \label{fig:N2HPHCOPcomp}
\end{center}
\end{figure}

It has been reported that N$_2$H$^+$/HCO$^+$ abundance decreases as a high mass clump evolves \citep{sanhueza2012chemistry,lee2003chemistry,bergin2007cold}. A comparison with the sources in \citet{zhang2016global} shows that MM sources have a larger median abundance ratio than $\HII$/PDR but lesser than protostellar.
We also find that the median abundance ratio of MM sources is higher than that of IR-bright souces in \citet{miettinen2014malt90}. As shown in Fig.~\ref{fig:N2HPHCOPcomp}, we find a higher median ratio for MM$_R$ and MM$_U$ sources compared to $\HII$ regions and PDRs in \citet{rathborne2016molecular} and \citet{urquhart2019atlasgal}. Given that the N$_2$H$^+$/HCO$^+$ intensity ratio declines as the source evolves, this is indicative of MM sources being 
less evolved than $\HII$ regions and PDRs. While, a KS test performed on  \citet{urquhart2019atlasgal} identifies maser sources, protostellar sources and $\HII$ regions as distinct samples, the same test performed on \citet{rathborne2016molecular} identifies maser sources and PDRs as distinct. Hence, \citet{rathborne2016molecular} suggests maser sources to be less evolved than PDRs and the N$_2$H$^+$/HCO$^+$ line intensity ratios of \citet{urquhart2019atlasgal} is  consistent with methanol maser sources being at a later evolutionary stage than protostellar phase and less evolved than $\HII$ regions.

\begin{figure}
\begin{center}
\includegraphics[width=0.4\textwidth, trim= 0 0.3cm 0 0]{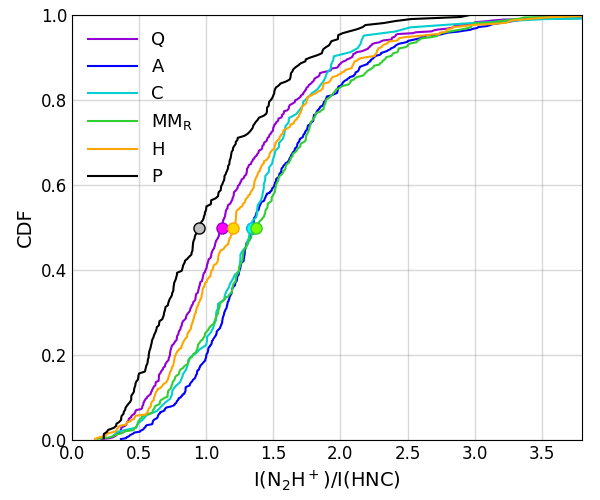}
\includegraphics[width=0.4\textwidth, trim= 0 0.4cm 0 0]{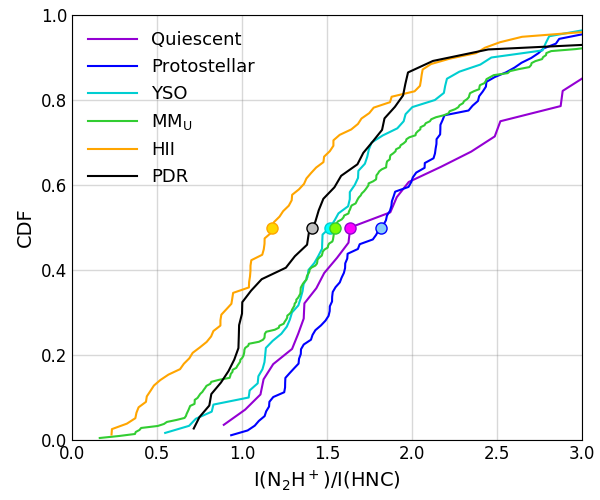}

\caption{MM sources compared with the N$_2$H$^+$/HNC integrated intensity ratios of \citet{rathborne2016molecular} and \citet{urquhart2019atlasgal} shown in top and bottom panels respectively. The filled circles on the CDFs represent the median line intensity ratios. Letters ‘Q’, 'A', 'C', 'H' and 'P' denotes 
quiescent, protostellar, compact  $\HII$ regions, extended  $\HII$ regions and  photo-dissociation regions respectively.} \label{fig:N2HPHNCcomp}
\end{center}
\end{figure}

Finally, we investigate how the N$_2$H$^+$/HNC ratio of MM sources compare with clumps in different evolutionary stages. The median column density and abundance ratios for MM sources are similar to IRDCs in \citet{liu2013systemic}. Fig.~\ref{fig:N2HPHNCcomp} shows the N$_2$H$^+$/HNC intensity ratio for sources in different evolutionary stages as per \citet{rathborne2016molecular} and \citet{urquhart2019atlasgal}. It can be seen that the N$_2$H$^+$/HNC ratio declines as the source evolves, which is contrary to what was observed by \citet{sanhueza2012chemistry}. We also see that the median line ratio of MM$_R$ and MM$_U$ sources is higher than that of $\HII$ regions and PDRs (these are supported by the KS tests), which again is consistent with methanol maser sources tracing an earlier evolutionary stage than $\HII$ regions. 

In short, the chemical properties are in agreement with the picture of methanol masers tracing an early evolutionary stage prior to the formation of PDRs and possibly $\HII$ regions (clearing phase). While the median line ratios of the sources with methanol masers are generally in between that of sources in the protostellar phase and sources identified as $\HII$ regions and PDRs, there is considerable overlap in the CDFs. This may be either due to the methanol maser phase overlapping with that of massive young stellar objects and $\HII$ regions, or due to chemical properties having lower sensitivity (compared to continuum studies) with respect to discriminating between different evolutionary phases.

\section{Conclusion}\label{sec5}
We have investigated the chemical environments of 6.7~GHz methanol maser hosts using N$_2$H$^+$(1-0), HCO$^+$(1-0), HCN(1-0) and HNC(1-0) molecular lines from the MALT90 survey, where the sources were selected to be representative of the full sample in Paper~I. The abundances of these molecular species is in congruence to the typical values observed in massive star forming regions. We do not find any correlation between molecular abundance and dust temperature, which suggest that the molecular abundances do not evolve much during the evolutionary stage traced by methanol masers. The HNC and HCO$^+$ integrated intensities, on the other hand, showed a weak positive correlation with source evolution. The HCN/HNC, HNC/HCO$^+$, N$_2$H$^+$/HCO$^+$ and N$_2$H$^+$/HNC ratios are in agreement with the picture of the methanol maser phase occurring prior to the formation of $\HII$ regions/PDRs. The analysis of 
HCN/HNC ratios hint methanol maser sources being more evolved than protostellar sources.These results are consistent with earlier studies including the SED studies of Paper~I wherein a majority of the methanol masers had mass to luminosity ratios suggestive of them being in accretion phase, with the masers turning off in the clearing phase. 

\section{Acknowledgements}
We thank the anonymous referee whose comments helped in improving the paper. This research has made use of NASA’s Astrophysics Data System and VIZIER service operated at CDS, Strasbourg, France. The \textsc{\textmd{matplotlib}} package \citep{hunter2007matplotlib} for python was used for making plots. We have used the online freemium academic writing environment Overleaf (\url{https://www.overleaf.com/}) for typesetting this paper.

\section{Data Availability}
The molecular line data can be accessed through the Australia Telescope Online Archive (ATOA) (\url{http://atoa.atnf.csiro.au/MALT90}). The PACS and SPIRE data is available at the ESA Hershel Science Archive (\url{http://archives.esac.esa.int/hsa/whsa/}). The ATLASGAL data can be obtained at \url{https://atlasgal.mpifr-bonn.mpg.de/cgi-bin/ATLASGAL_DATASETS.cgi}.

\bibliographystyle{mnras}
\bibliography{ref}

\appendix

\section{Chemistry of molecules}
\subsection{HCN,HNC (hydrogen (iso)-cyanide)} \label{chemistryHCN}
HCN and its metastable geometrical isomer HNC are ubiquitous in the dense star forming regions and are considered to be probing high-density gas. They are also good tracers of infall motions in star forming regions \citep{wu2003indications}. According to gas-phase chemical models, HCN and HNC are primarily produced through the following dissociative recombination reaction of HCNH$^+$ \citep{mendes2012cold}:
\begin{equation}
 \mathrm{HCNH^+ + e^{-}}\longrightarrow\begin{cases} \label{eq:HCNform}
    \mathrm{HCN+H} \\
   \mathrm{HNC+H}
  \end{cases}
\end{equation}
Here both HCN and HNC are formed in equal measures causing HCN/HNC abundance ratio close to unity ($\simeq 0.9$). An additional channel where only HNC is produced, is through the dissociative recombination of H$_2$CN$^+$ and H$_2$NC$^+$ \citep{allen1980possible}:
\begin{equation}
\begin{rcases}
\mathrm{H_2CN^+ + e^{-}}\\
\mathrm{H_2NC^+ + e^{-}}\\
\end{rcases}
\longrightarrow \mathrm{HNC + H}
\end{equation}
This results HCN/HNC ratio to slightly rise above unity. These molecules are also formed through other channels say,
\begin{equation}
\mathrm{CH_2 + N} \longrightarrow \mathrm{HCN + H}
\end{equation}
\begin{equation}
\mathrm{NH_2 + C \longrightarrow HNC + H}
\end{equation}
These  are then followed by rapid isomerisation reactions, once again resulting in near unity HCN/HNC ratio. 
The destruction of HCN and HNC molecules in dense clouds, primarily happens via the following reaction channels,
\begin{equation}
\mathrm{HCN + H^+ \longrightarrow HCN^+ + H} \\
\end{equation}
\begin{equation}
\mathrm{HCN + HCO^+ \longrightarrow H_2CN^+ + CO} \\
\end{equation}
\begin{equation}
\mathrm{HNC + H \longrightarrow HCN + H }\\ \label{eq:HNCconsum}
\end{equation}
\begin{equation}
\mathrm{HNC + H^+ \longrightarrow HCN + H^+} \\
\end{equation}
\begin{equation}
\mathrm{HNC + O \longrightarrow NH + CO} 
\end{equation} 
\subsection{HCO$^+$} \label{chemistryHCOP}
HCO$^+$ molecule, regarded as one of the excellent tracers of high-density gas, is mainly formed via the following two channels  \citep{schlingman2011bolocam} :
\begin{equation}
\mathrm{H_3^+ + CO \longrightarrow HCO^+ + H_2 }
\end{equation}
\begin{equation}
\mathrm{N_{2}H^+ + CO \longrightarrow HCO^+ + N_2}
\end{equation}
In the cold dense molecular clouds ($n$> 10$^{4}$ cm$^{-3}$, T$_{kin}$< 24~K), CO is frozen out on to the dust grains, resulting in the depletion of HCO$^+$. As the temperature increases, CO is liberated from the dust grains, enhancing the production of HCO$^+$. Hence, the column densities and abundances of HCO$^+$ molecule is expected to increase as the source evolves. The destruction of HCO$^+$ is primarily due to the dissociative recombination reaction:
\begin{equation*}
\mathrm{HCO^+ + e^{-} \longrightarrow CO + H}
\end{equation*}

\subsection{N$_2$H$^{+}$}\label{chemistryN2HP}
 These molecules are primarily formed through the gas-phase reaction :
\begin{equation}
   \mathrm{ H_3^+ + N_2 \longrightarrow N_2H^+ + H_2}
\end{equation}
and destroyed by the reaction with CO molecules in gas phase producing HCO$^+$ \citet{bergin1997chemical}
\begin{equation}
    \mathrm{N_2H^+ + CO \longrightarrow HCO^+ + N_2}
\end{equation}

Free electrons can also destroy N$_2$H$^+$ molecules via the dissociative recombination reaction and it's the primary destruction mechanism when CO is depleted.
\begin{equation}
   \mathrm{ H_3^+ + N_2 \longrightarrow N_2H^+ + H_2}
\end{equation}
\begin{equation}
    \mathrm{N_2H^+ + e^- \longrightarrow N_2 + H/NH + H}
\end{equation}

\section{Comparison of properties of 68 MM sources with that of larger MMB sample}\label{comp68}
\begin{figure}
\begin{center}
\includegraphics[width=0.4\textwidth, trim= 0 0.3cm 0 0]{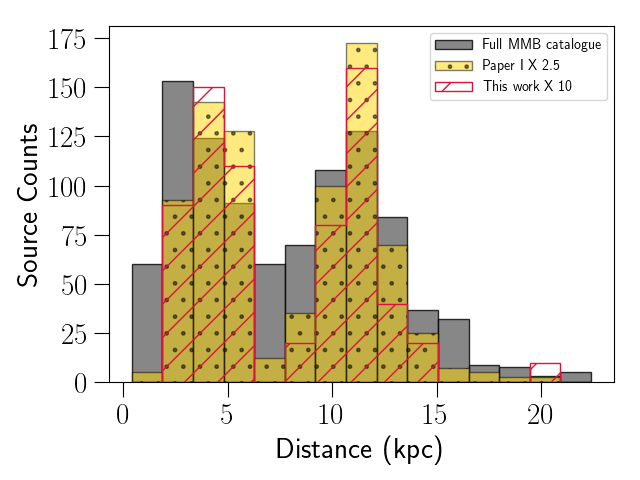}
\includegraphics[width=0.4\textwidth, trim= 0 0.3cm 0 0]{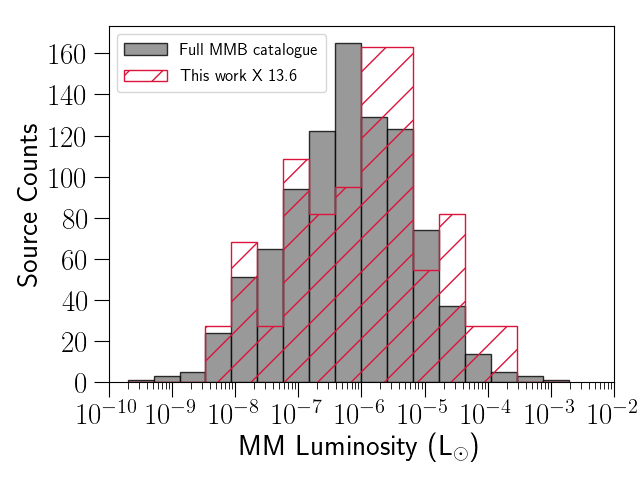}
\includegraphics[width=0.4\textwidth, trim= 0 0.3cm 0 0]{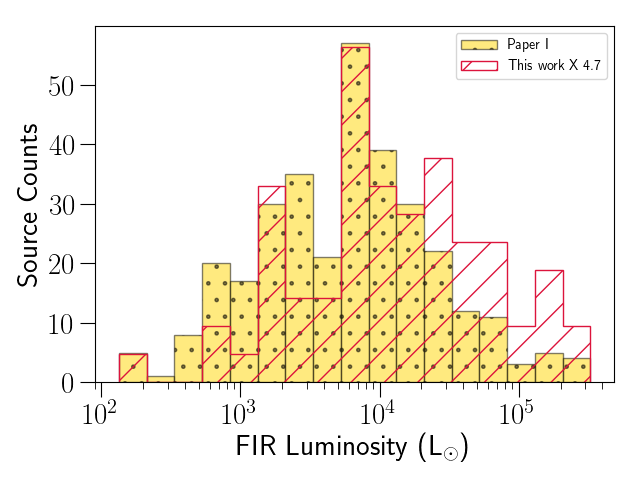}
\caption{The top, middle and bottom panels show the histogram for distances, methanol maser luminosities and FIR luminosities towards the 68 sources of this paper (red with hatches), 321 sources of Paper~I (yellow with dots) and the full MMB catalogue (grey). The histograms have been scaled for ease of comparison, with the scale factors being indicated in the legend of each panel.} \label{compfig68}
\end{center}
\end{figure}

Since the goal of our work is to examine the chemical properties of sources that host 6.7~GHz methanol masers, it is important to ensure that our sample is representative of the Galactic population. To this effect, we have examined the distances, methanol maser luminosities and the FIR luminosity of the host sources of our sample and compared them with that of the full MMB catalogue/sources in Paper~I~(Fig.~\ref{compfig68}). 
While the distribution of distances of our sample is similar to that of the full MMB catalogue and the sample studied in Paper~I, there appears to be a slight bias towards larger maser luminosities and FIR luminosities in this study. In order to check for any bias quantitatively, the Student's T-test was performed on the samples. The $p$-value obtained was 0.92 for the methanol maser distribution and 0.2 for the FIR luminosity distribution. These are significantly higher than 0.05, which suggests that the differences between the maser populations are not statistically significant.

\section{Comparison of dust temperatures of 68 MM sources with that of previous works}\label{temp68}
\begin{figure}
\begin{center}
\includegraphics[width=0.4\textwidth, trim= 0 0.3cm 0 0]{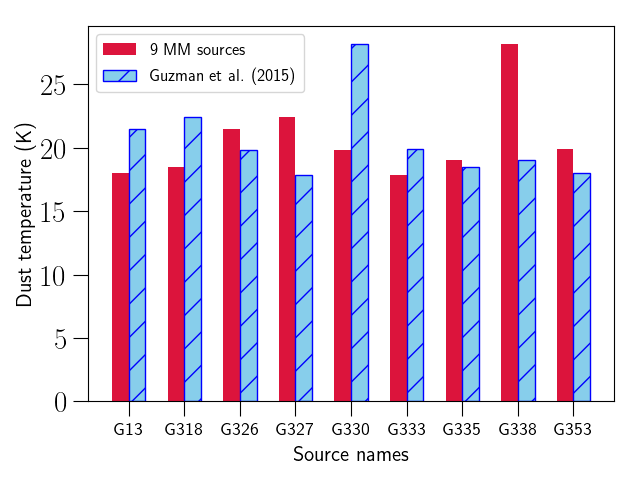}
\includegraphics[width=0.4\textwidth, trim= 0 0.3cm 0 0]{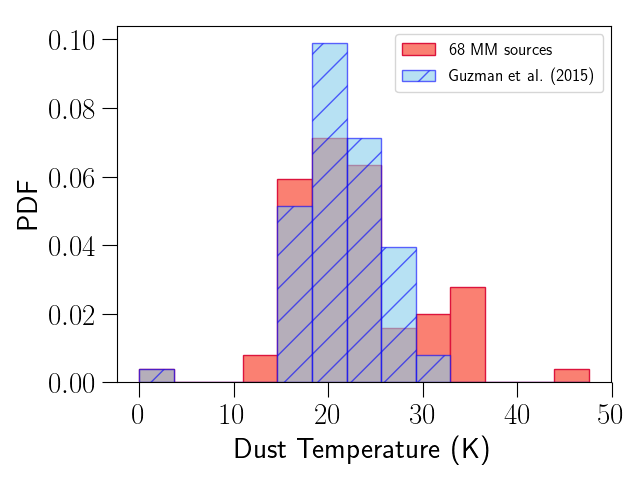}
\caption{Average dust temperatures of sources obtained from pixel-by-pixel fitting of the brightest pixels, compared with the average dust temperatures reported by \citet{guzman2015far}, is shown in the left panel. The right panel shows the PDFs of dust temperatures of 68 sources (taken from Paper~I) and the average dust temperatures given in \citet{guzman2015far}.} \label{comptempfig}
\end{center}
\end{figure}
The left panel of Fig.~\ref{comptempfig} illustrates the average dust temperature calculated for all the bright pixels (S/N $\ge5$) of each of the nine source (chosen for pixel-by-pixel study) and average dust temperature calculated for the same sources by \citep{guzman2015far}. Except for two sources, the dust temperature values of the nine sources agree well with what is reported by \citet{guzman2015far}. The right panel of Fig.~\ref{comptempfig} shows the PDFs of dust temperatures of 68 MM sources (taken from Paper~I) with that of \citet{guzman2015far}. The distributions can be seen comparable with each other.

\bsp	
\label{lastpage}
\end{document}